\documentclass[apj]{emulateapj}

\usepackage{mathptmx}
\usepackage{color}
\usepackage{subfigure}

\journalinfo{Draft version \today}
\slugcomment{Accepted for publication in ApJ}
\shorttitle{PHL 1811 ANALOGS AT HIGH REDSHIFT}
\shortauthors{WU ET AL.}

\def\simgt{\lower 2pt \hbox{$\, \buildrel {\scriptstyle >}\over {\scriptstyle \sim}\,$}}
\def\simlt{\lower 2pt \hbox{$\, \buildrel {\scriptstyle <}\over {\scriptstyle \sim}\,$}}
\def\chandra{{\it Chandra}}

\def\xmm{\hbox{\it XMM-Newton}}

\def\pl{\hbox{power-law}}

\def\aox{$\alpha_{\rm ox}$}
\def\daox{$\Delta\alpha_{\rm ox}$}

\def\xray{\hbox{X-ray}}
\def\phl{PHL~1811}

\begin{document}


\title{A Population of X-ray Weak Quasars: PHL~1811 Analogs at High Redshift}


\author{Jianfeng~Wu\altaffilmark{1,2}, 
W.~N.~Brandt\altaffilmark{1,2},
Patrick~B.~Hall\altaffilmark{3},
Robert~R.~Gibson\altaffilmark{4},
Gordon~T.~Richards\altaffilmark{5},
Donald~P.~Schneider\altaffilmark{1},
Ohad~Shemmer\altaffilmark{6},
Dennis~W.~Just\altaffilmark{7},
Sarah~J.~Schmidt\altaffilmark{4}
}

\altaffiltext{1}
    {Department of Astronomy \& Astrophysics, The Pennsylvania State
     University, 525 Davey Lab, University Park, PA 16802, USA}
\altaffiltext{2}
    {Institute for Gravitation and the Cosmos, The Pennsylvania State 
     University, University Park, PA 16802, USA}
\altaffiltext{3}
    {Department of Physics \& Astronomy, York University, 4700 Keele Street, 
     Toronto, ON M3J 1P3, Canada}
\altaffiltext{4}
    {Department of Astronomy, University of Washington, Box 351580, 
     Seattle, WA 98195, USA}
\altaffiltext{5}
    {Department of Physics, Drexel University, 3141 Chestnut Street, 
     Philadelphia, PA 19104, USA}
\altaffiltext{6}
    {Department of Physics, University of North Texas, Denton, TX 76203, USA}
\altaffiltext{7}
    {Steward Observatory, University of Arizona, 933 North Cherry Avenue, Tucson, 
     AZ 85721, USA}
\email{jfwu@astro.psu.edu}


\begin{abstract}
We report the results from \chandra\ and \xmm\ observations of
a sample of 10 type~1 quasars selected to have unusual UV emission-line
properties (weak and blueshifted high-ionization lines; 
strong UV Fe emission) similar to those of \phl, a confirmed intrinsically
\xray\ weak quasar. These quasars were identified by the Sloan 
Digital Sky Survey at high redshift ($z\approx2.2$); eight are radio quiet
while two are radio intermediate. All of the \hbox{radio-quiet}
\phl\ analogs, without exception, are notably \xray\ weak by a mean factor 
of $\approx13$. These sources lack broad absorption lines 
and have blue UV/optical continua, supporting the hypothesis that they are 
intrinsically \xray\ weak like \phl\ itself. However, their 
average \xray\ spectrum appears to be harder than those of typical 
quasars, which may indicate the presence of heavy intrinsic \xray\ 
absorption. Our sample of \hbox{radio-quiet}
\phl\ analogs supports a connection between an \xray\
weak spectral energy distribution (SED) and \phl-like UV 
emission lines; this connection provides an economical way to identify \xray\ 
weak type~1 quasars. The fraction of \hbox{radio-quiet} \phl\ analogs in the
\hbox{radio-quiet} quasar population is estimated
to be $\lesssim1.2\%$. We have investigated correlations between
relative \xray\ brightness and UV emission-line properties 
(e.g., C~{\sc iv} equivalent width and blueshift) for a sample combining
our \hbox{radio-quiet} \phl\ analogs, \phl\ itself, and typical 
type~1 quasars. These correlation analyses suggest that \phl\
analogs may have extreme wind-dominated broad emission-line
regions. 
Observationally, the \hbox{radio-quiet} \phl\ analogs appear to be a 
subset ($\approx30\%$) of \hbox{radio-quiet} \hbox{weak-line} quasars. 
The existence of a subset of quasars in which high-ionization ``shielding gas''
covers most of the BELR, but little more than the BELR, 
could potentially unify the \phl\ analogs and WLQs.
The two \hbox{radio-intermediate} \phl\ analogs are \xray\ bright. \xray\
spectral analyses and consideration of their multiwavelength properties suggest
that one of them has jet-dominated \xray\ emission, while 
the nature of the other remains unclear. 
\end{abstract}

\keywords{galaxies: active --- galaxies: nuclei ---
quasars: emission lines --- \hbox{X-rays}: galaxies}


\section{Introduction}\label{intro}

A central tenet of \hbox{X-ray} astronomy is that luminous \hbox{X-ray} emission 
is a universal property of efficiently accreting supermassive black holes (SMBHs). 
This idea underlies the utility of extragalactic \hbox{X-ray} surveys for finding 
active galactic nuclei (AGNs) throughout the Universe. While this tenet has generally 
withstood observational tests (e.g., Avni \& Tananbaum 1986; Mushotzky 2004; 
Brandt \& Hasinger 2005; Gibson et~al. 2008a; Brandt \& Alexander 2010; 
and references therein), it is poorly understood physically. The accretion-disk 
corona (ADC), putatively 
responsible for creating most of the observed AGN \hbox{X-ray} emission via Compton
upscattering of lower energy disk photons, is still not well understood, and its strong level 
of \hbox{X-ray} emission cannot yet be reliably derived from {\it ab initio\/} physics 
(e.g., Galeev et~al. 1979; Fabian et~al.\ 2000; Miller \& Stone 2000; 
Krolik 2007). 

Notably, there are a few examples of AGNs where the ADC appears to 
emit \hbox{X-rays} much more weakly, by a factor of \hbox{$\approx 10$--100}, 
than expected based upon the emission at longer wavelengths. These
strange objects, which persist in showing spectral energy
distributions (SEDs) entirely out of keeping with their luminosity, may ultimately teach 
us more than a host which radiate according to rule (cf. Eddington 1922). 
The best-studied such case is the \hbox{radio-quiet} quasar \phl\ 
($z=0.19$; Leighly et~al. 2007ab). In multiple \hbox{X-ray} observations, this 
quasar has been found to be consistently \hbox{X-ray} weak relative to expectations 
from the \hbox{\aox-$L_{2500~\mathring{\rm{A}}}$} relation by a factor of 
\hbox{$\approx 30$--100}.\footnote{\aox\ is defined to be the 
slope of a power law connecting the rest-frame 2500~\AA\ and 2~keV 
monochromatic luminosities; i.e., 
$\alpha_{\rm ox}=0.3838 \log(L_{\rm 2~keV}/L_{2500~\mathring{\rm{A}}})$.
This quantity is well known to be correlated with $L_{2500~\mathring{\rm{A}}}$ 
(e.g., Steffen et~al. 2006 and references therein).} Its \hbox{X-ray} spectrum 
shows no evidence for absorption of an intrinsically strong underlying \hbox{X-ray} 
continuum, and large-amplitude variability demonstrates that the \hbox{X-rays} 
are not strongly scattered in the nuclear region. \phl\ thus appears to be 
{\it intrinsically\/} \hbox{X-ray} weak. The ultimate physical reason for this 
intrinsic \hbox{X-ray} weakness is poorly understood. It may be due to a high 
accretion rate (i.e., $L/L_{\rm Edd}$) onto the SMBH; this property could quench or 
catastrophically cool the ADC, perhaps due to ``trapping'' effects 
(e.g., Begelman 1978). 

The UV/optical spectrum of \phl\ is also unusual 
(Leighly et~al. 2007b). It shows no clear forbidden or semi-forbidden 
line emission. The C~{\sc iv} $\lambda 1549$ emission line is weak by a 
factor of $\approx 5$ compared to composite quasar spectra 
(only $\approx$1\% of SDSS quasars with similar luminosity have
such weak \ion{C}{4} lines; e.g., Shen et~al. 2010). The \ion{C}{4} line is
also blueshifted and asymmetric; the observed \ion{C}{4} characteristics may be indicating
the presence of a strong radiatively driven wind that
dominates the broad emission-line region (BELR; e.g., Richards et~al. 2010). 
The \hbox{near-UV} spectrum is dominated by 
strong Fe~{\sc ii} and Fe~{\sc iii} emission lines, and unusual \hbox{low-ionization} 
emission lines such as Na~{\sc i} D and Ca~{\sc ii} H and K are observed. 
Based on photoionization modeling, Leighly et~al.\ (2007b) suggest that the 
unusual UV/optical emission-line properties of \phl\ are due to a weak 
ionizing continuum, inferred from its soft (i.e., \hbox{X-ray} weak and 
UV/optical strong) SED. This apparent connection between the SED and 
emission-line properties does not, unfortunately, directly clarify 
the ultimate physical cause of the intrinsic \hbox{X-ray} weakness. 
However, if correct, this suggestion should provide an economical method of 
identifying additional examples of intrinsically \hbox{X-ray} weak quasars from large 
optical spectroscopic databases. Investigations of these additional examples
might then provide broader context and insights into the cause of the intrinsic \hbox{X-ray} 
weakness. Additional evidence for a connection between a soft SED 
and \phl-like emission-line properties comes from the finding that the 
extremely luminous quasar SDSS~J1521+5202 ($z=2.19$; $M_i=-30.19$) is also remarkably \hbox{X-ray} 
weak (by a factor of $\approx30$; Just et~al. 2007) and has similar UV 
emission-line properties to \phl. However, the sample size remains 
too small to claim an empirical connection reliably. 

One other recent result that is relevant to the issue of intrinsically 
\hbox{X-ray} weak quasars is the finding by Miniutti et~al. (2009) of an
abrupt and remarkable drop by a factor of $\approx 200$ in the 
\hbox{X-ray} luminosity of the well-studied quasar PHL~1092 
($z=0.40$). This quasar appears to have made a strong transition 
between \hbox{X-ray} bright and \hbox{X-ray} weak states, 
perhaps associated with evolving instabilities in its ADC
(see Miniutti et~al. 2009; Sobolewska et~al. 2009). 
Interestingly, PHL~1092 was noted to have similar emission-line 
properties to \phl\ prior to the discovery of its \hbox{X-ray} 
luminosity drop (Leighly et~al. 2007b). Somewhat surprisingly, the wavelength
region from \ion{Mg}{2} to H$\beta$ in 
PHL~1092 did not show dramatic changes associated with its transition from 
a normal SED to an \xray\ weak SED (Miniutti et ~al. 2009). 

Given the potential physical importance of quasars with \phl-like 
emission-line properties (hereafter \phl\ analogs), we proposed to use \chandra\ to enlarge
substantially the sample of these objects with sensitive \hbox{X-ray} 
coverage. Our observation targets were selected from the vast
Sloan Digital Sky Survey (SDSS; York et~al. 2000) spectroscopic database 
to be sufficiently optically bright to allow derivation of 
tight optical-to-\hbox{X-ray} SED constraints with 
short \hbox{(5--13~ks)} \chandra\ observations. We also searched for
and utilized sensitive archival \hbox{X-ray} coverage of a few \phl\ analogs. 
The closely related goals of our project were the following: 
(1) To determine if there is evidence for a reasonably sized {\it population\/} of 
intrinsically \hbox{X-ray} weak quasars. If such a population indeed 
exists, it would present an interesting challenge to the universality 
of luminous 
\hbox{X-ray} emission from quasars, and one that might lead to 
insights into when luminous ADCs do and do not form; 
(2) To assess empirically if there is a physical connection between 
soft (i.e., \hbox{X-ray} weak and UV/optical strong) quasar SEDs and 
\phl-like emission-line properties, as has been proposed to
be the case from photoionization modeling;  
(3) To investigate if selection upon \phl-like emission-line 
properties provides a practical and economical way of finding \hbox{X-ray} 
weak quasars at high redshift; and 
(4) To provoke further \hbox{X-ray} studies of \phl\ analogs. For
example, \hbox{X-ray} spectroscopy of any \hbox{X-ray} weak members
of this class should be able to test directly if they, like \phl\
itself, are intrinsically \hbox{X-ray} weak rather than absorbed, and
it should also provide estimates of their $L/L_{\rm Edd}$ values. Consistent 
\hbox{X-ray} monitoring over long timescales might also reveal strong X-ray
variability (perhaps associated with ADC instabilities), as has been 
found for PHL~1092. 

In \S\ref{sample} we describe the selection of our sample of \phl\ analogs 
and measurements of their rest-frame UV spectral properties. 
In \S\ref{xray} we describe the relevant \hbox{X-ray} data analyses. Overall 
results and associated discussion are presented in \S\ref{discuss}. 
Throughout this paper, we adopt a cosmology with
$H_0=70.5$~km~s$^{-1}$~Mpc$^{-1}$, 
$\Omega_{\rm M}=0.274$, and 
$\Omega_{\Lambda}=0.726$
(e.g., Komatsu et~al. 2009). 
\begin{figure}[t]
    \centering
    \includegraphics[width=3.5in]{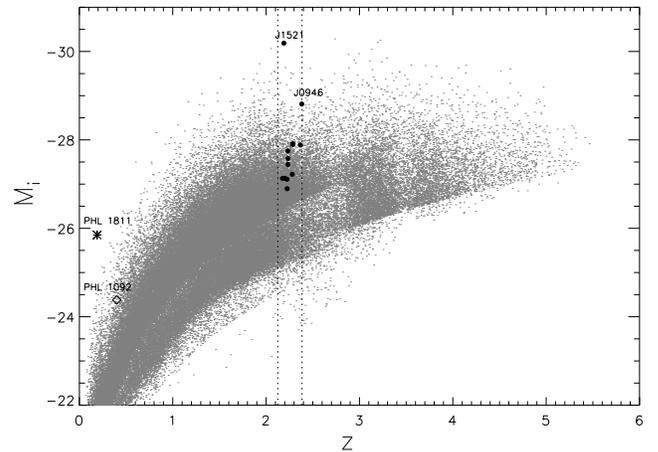}
    \caption{\footnotesize{SDSS absolute $i$-band magnitude, $M_i$, plotted versus redshift, $z$. The filled
              black circles show our sample of 11 selected quasars, the asterisk represents \phl, the open 
              diamond represents PHL~1092, and the small grey dots represent the 105,783 objects in the SDSS 
              DR7 quasar catalog (Schneider et~al. 2010). The two most luminous sources in our sample are 
              labelled in the format of 'J$hhmm$' for brevity. The vertical dotted lines
              show the redshift criterion for our sample selection ($2.125\leq z \leq 2.385$).}
             \label{zMi_fig}}
\end{figure}


\section{Sample Selection and Rest-Frame UV Spectral Measurements}\label{sample}

\subsection{Selection of \phl\ Analogs}\label{sample:select}

\begin{figure*}[t]
    \centering
    \includegraphics[width=6.3in]{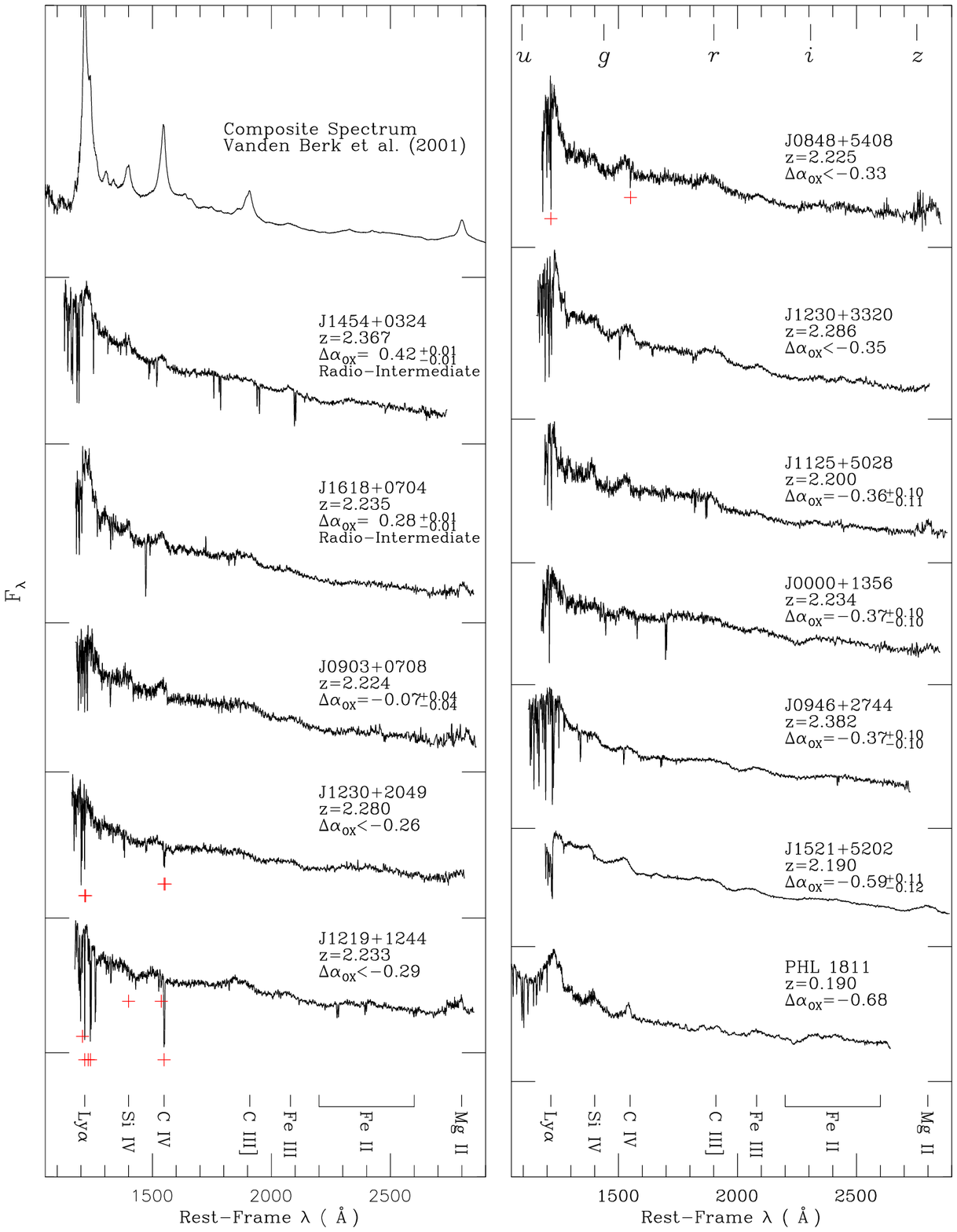}
    \caption{\footnotesize{SDSS spectra for the 11 sources in our sample of potential \phl\ analogs 
             (J0903+0708 is later excluded from our sample, see \S\ref{sample:0903}), ordered by
             \daox\ (see \S\ref{xray} for definition). The \daox\ values and their 
             error bars (if the source is detected in \hbox{X-rays}) are shown for each source. 
             The name of each source is labeled in the format of 'J{\it hhmm+ddmm}'.
             The \hbox{$y$-coordinates} are the flux density ($F_\lambda$) in arbitrary linear units. 
             The tick marks on the \hbox{$y$-axis} show the zero flux density level for each normalized
             spectrum. The spectra have been smoothed using a \hbox{5-pixel} \hbox{sliding-box} filter. 
             Emission lines, including Ly$\alpha\ \lambda$1216, Si~{\sc iv}~$\lambda$1400, C~{\sc iv}~$\lambda$1549, 
             Fe~{\sc iii}~$\lambda$2080, Fe~{\sc ii}~$\lambda$2200-2600, and Mg~{\sc ii}~$\lambda$2799 are labeled 
             in each panel. The associated narrow absorption lines are 
             labeled with red `+' signs (the two associated NAL systems
             of J1230+2049 overlap each other because of their very close redshifts; see Table~\ref{abs_table}). 
             The rest-frame effective wavelengths of the SDSS $ugriz$ bands for the median redshift of our \phl\ analogs 
             ($z=2.23$) are labelled at the top of the right panel. The spectral resolution 
             is $R\approx$~2000. Also included are the 
             composite spectrum of SDSS quasars by Vanden~Berk et~al. (2001) and the \phl\ spectrum from the 
             {\it Hubble Space Telescope} (Leighly et~al. 2007b).}
             \label{spec_fig}}
\end{figure*}

We first compiled a sample of the 1621 objects classified as \verb+QSO+ or \verb+HIZ_QSO+ 
in the SDSS Data Release 7 (DR7; Abazajian et~al. 2009) Catalog Archive Server (CAS)
with $m_r \leq 18.8$ and 
$2.125 \leq z \leq 2.385$ (see Fig.~\ref{zMi_fig}). The magnitude limit selects objects
that are sufficiently bright
for short \chandra\ observations to provide tight constraints on their optical-to-\xray\ 
SED properties. The selected redshift range provides the best possible coverage between 
Ly$\alpha$ and \ion{Fe}{2} (and usually \ion{Mg}{2}) in the SDSS spectra, which allows reliable 
identification of \phl\ analogs. These analogs were systematically selected as
quasars having weak and blueshifted \hbox{high-ionization} lines, such as \ion{C}{4}~$\lambda1549$ and
\ion{Si}{4}~$\lambda1400$,\footnote{This line is, in fact, a blend of Si~{\sc iv} and O~{\sc iv}]; 
we refer to it simply as Si~{\sc iv} for convenience.} as well as strong UV \ion{Fe}{2} 
($2200$--$2600$~\AA) and \ion{Fe}{3} UV48 (2080~\AA) emission. 
The criteria upon rest-frame equivalent widths ($W_r$) of 
\hbox{high-ionization} lines extracted from the SDSS CAS \verb+SpecObj+ table 
were $W_r$(\ion{Si}{4}) $\leq 15$~\AA, $-15$~\AA $\leq W_r$(\ion{C}{4}) $\leq 20$~\AA, 
and $\chi^2\leq 20$ for the \ion{C}{4} line fit; the $\chi^2$ fit-quality criterion ensures 
reliable measurements of \ion{C}{4} line properties.\footnote{We also required
$W_r({\rm Ly}\alpha)\leq 105$~\AA\ and $W_r({\rm N~{\sc v}}~\lambda1240)\leq 85$~\AA\ to
eliminate quasars with strong Ly$\alpha$+N~{\sc v} blends.}
All 54 spectra meeting these criteria were then inspected visually by
P.B.H. We first excluded sources that are misclassified stars or are otherwise spurious, 
as well as sources that have bad redshift measurements or detectable UV broad absorption lines (BALs). 
BAL quasars were excluded because they typically show substantial \xray\
absorption (e.g., Gallagher et~al. 2002, 2006; Gibson et~al. 2009a).
After this examination we had 32 sources which are \hbox{non-BAL} \hbox{weak-line} quasars (WLQs; see
\S\ref{discuss:wlq} for further details). 
Among these 32 sources, objects with strong UV \ion{Fe}{2} and \ion{Fe}{3} UV48 emission and whose
high-ionization lines showed a strong blueshifted component (with the magnitude of the blueshift
exceeding 800~km~s$^{-1}$) were kept in the sample. We rejected 13 objects primarily due to weak Fe emission,
and 8 objects primarily due to a low or unknown blueshift.
After doing this, 11 quasars remain in our sample.\footnote{One radio-quiet source, J0903+0708, 
was later removed from the sample based upon additional observations (see \S\ref{sample:0903} for details).}
Fig.~\ref{spec_fig} shows the SDSS spectra for 
these potential \phl\ analogs. The SDSS quasar composite spectrum from Vanden~Berk
et~al. (2001) and the spectrum for \phl\ from the {\it Hubble Space Telescope} (Leighly et~al. 2007b) 
are also included in
Fig.~\ref{spec_fig}. Nine of the 11 objects are \hbox{radio-quiet} ($R<10$; see \S\ref{xray} for
the definition of $R$). J1454+0324 and J1618+0704 are 
\hbox{radio-intermediate} with $R=12.8$ and $R=35.0$, respectively. Three of the 11 objects
already had sensitive archival \xray\ coverage (J1219+1244 and J1521+5202 by \chandra, J1230+2049 by \xmm). 
We proposed short (5--13 ks) 
\chandra\ observations for the other eight objects, and were awarded observing time 
in Cycle 11 (see Table~\ref{log_table} for an \xray\ observation log). 

\subsection{Measurements of Redshifts and Rest-Frame UV Emission-line Properties}\label{sample:line}

The redshifts and their uncertainties (see Table~\ref{abs_table}) for our \phl\ analogs are the \verb+spectro1d+ 
redshifts from the SDSS CAS in most cases. These redshifts 
were obtained by fitting the peaks in the SDSS spectra with a set of strong quasar
emission lines (see \S4.2.10.1 of Stoughton et~al. 2002 for details). The SDSS CAS redshift measurements may not 
be precise for our sample because of the weak, blueshifted, and asymmetric emission lines. We therefore 
examined the redshift values for our \phl\ analogs carefully. For four quasars 
whose CAS redshifts were suspect based on this examination, we adopt redshifts and uncertainties from 
other approaches: for J1125+4028, the redshift of the longer-wavelength peak of the 
double-peaked \ion{Mg}{2} emission is adopted as systemic; for both J1219+1244 and J1230+2049, 
we adopt the redshift of a strong (but not broad) absorption system; 
for J1521+5202, we follow Just et~al.~(2007) and adopt $z=2.19$ based on \ion{Mg}{2} emission. The differences
between these updated redshift values and the CAS values have a magnitude of $\Delta z=0.02$--$0.05$.  
The final redshift values adopted are listed as $z_q$ in Table~\ref{abs_table}. 
Redshifts of narrow absorption-line systems (NALs) are also listed in Table~\ref{abs_table}. Three \phl\
analogs (J0848+5408, J1219+1244, and J1230+2049) have associated NALs (defined as $|v|<5000$~km~s$^{-1}$), which
are labeled in Fig.~\ref{spec_fig} and in Table~\ref{abs_table}. The other NALs with $v<-5000$~km~s$^{-1}$ 
are most likely intervening absorption.

Hewett \& Wild (2010) reported improved redshift measurements with lower systematic uncertainties 
for SDSS quasars. Their measurements for our candidate \phl\ analogs are also included in Table~\ref{abs_table}, 
denoted as $z_{HW}$. These redshifts are either consistent with or slightly 
lower than $z_q$. However, the $z_{HW}$ values were determined via cross-correlation with quasar spectral 
templates. Given the unusual spectral properties of our candidate \phl\ analogs, 
we have not adopted the $z_{HW}$ values. Nevertheless, we will test whether adopting these different 
redshift values would affect our main conclusions in the following sections, particularly regarding
\ion{C}{4} blueshifts.

We measured emission-line properties interactively for \ion{C}{4}, \ion{Si}{4}, the $\lambda1900$ 
complex\footnote{Mainly C~{\sc iii}]~$\lambda1909$, but also including other features; see Note 
(b) of Table~\ref{qso_table}.}, and \ion{Fe}{3} UV48 (Table~\ref{qso_table}), 
since the automatic measurements stored in the CAS may be unreliable because of the weak, 
highly blueshifted lines of \phl\ analogs. We first fitted the continuum for 
each source following the method of Vanden~Berk et~al. (2001). The SDSS spectra were 
smoothed with a 5-pixel sliding-box filter. Regions of strong narrow absorption 
were interpolated across manually.
The fitting region for each emission line was then defined between lower and 
upper wavelength limits $\lambda_{lo}$ and $\lambda_{hi}$ (see Table~2 of Vanden~Berk et~al. 2001).
A \pl\ local continuum was fit to the lower and upper 10\%
of the wavelength region between $\lambda_{lo}$ and $\lambda_{hi}$.
After subtracting the continuum, we measured $W_r$, FWHM, the line 
dispersion ($\sigma_{\rm line}$, the second moment relative
to the flux-weighted mean wavelength of the line) and the associated uncertainties 
for each line. Blueshifts were calculated between the lab wavelength of a line in the
quasar rest frame (see Table~2 of Vanden~Berk et~al. 2001) and the observed mode of 
all pixels with heights greater than 50\% of the peak height of this line, 
where mode = 3$\times$median$-2\times$mean. In addition to the \ion{C}{4} blueshift errors quoted in 
Table~\ref{qso_table}, the uncertainties in the redshift measurements also produce further 
uncertainties for the \ion{C}{4} blueshift.
We also include corresponding measurements of the spectrum of \phl\ (Leighly et~al. 2007b), the
composite quasar spectrum of Vanden~Berk et~al. (2001), the `B4' composite spectrum of 
high-redshift ($2.06<z<3.33$), high-luminosity ($-28<M_i<-26$) quasars of Yip et~al.
(2004), and the spectrum of PHL~1092 (Leighly et~al. 2007b) in Table~\ref{qso_table} for comparison. 

Using our refined $W_r$ values, all our targets and \phl\ have 
$W_r$(\ion{C}{4}) $<$~11~\AA, $W_r$(\ion{Si}{4}) $<$~8~\AA, 
$3.5$~\AA~$< W_r$($\lambda$1900~\AA) $< 12.5$~\AA, and $2.5$~\AA~$< W_r$(\ion{Fe}{3}) $< 8$~\AA.
Our targets have significantly weaker \ion{C}{4} and \ion{Si}{4} lines than the required selection criteria.
For \phl\ analogs, the first three of these lines have average 
$W_r$ values which are only $\approx40\%$ of the average for normal quasars of similar
redshift and luminosity. Therefore, the sources in our sample have weak high-ionization lines and semi-forbidden lines
like \phl. In contrast, \ion{Fe}{3} UV48 has a 50\% larger average $W_r$
in \phl\ analogs. As compared to \phl\ itself, our \phl\ analogs have lines which 
are as weak but are typically broader (perhaps partly because our sources are more luminous and thus
likely have SMBHs of higher mass; e.g., Laor 2000) and more blueshifted. 
\begin{figure*}[ht]
    \centering
    \includegraphics[width=6.0in]{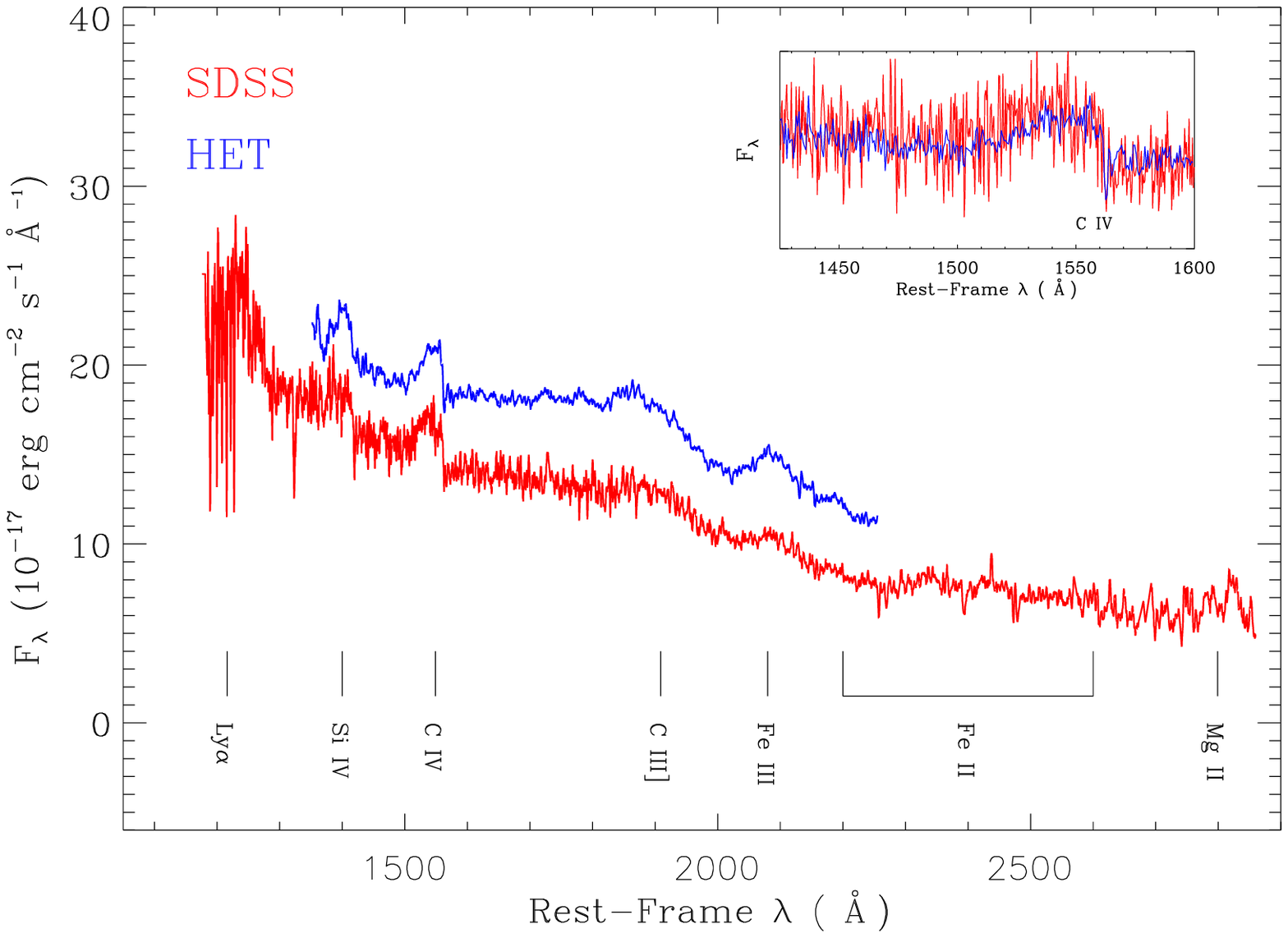}
    \caption{\footnotesize{Comparison of the coadded HET spectrum (blue) to the SDSS spectrum (red) of J0903+0708. 
             The \hbox{$y$-coordinates} are the flux density ($F_\lambda$). The HET spectrum has large absolute 
             flux uncertainty. The spectra have been smoothed using a \hbox{5-pixel} \hbox{sliding-box} filter. 
             The emission lines are labelled as in Fig.~\ref{spec_fig}.
             The inset in the upper right part of the figure shows a \hbox{zoom-in} of the C~{\sc iv} region without 
             smoothing. The two spectra in the inset are normalized according to their fluxes in the wavelength 
             range 1425--1450~\AA. The resolution of the SDSS spectrum is $R\approx$~2000, while
             that of the HET spectrum is $R\approx$~1000.}
             \label{het_fig}}
\end{figure*}

Some key emission lines in the rest-frame optical band, e.g., H$\alpha$, H$\beta$, and [\ion{O}{3}]~$\lambda5007$, are
not covered by the SDSS spectra of our \phl\ analogs due to their high redshifts. \phl\ has a narrow-line type 1 
optical spectrum as well as very weak [\ion{O}{3}] emission compared to typical type 1 
quasars (Leighly et~al. 2007b). Near-infrared spectroscopy of our \phl\ analogs is needed to measure the
properties of these emission lines. We report near-infrared spectroscopy of our most luminous \phl\ analog, J1521+5202,
in the Appendix of this work.

\subsection{J0903+0708: Not a \phl\ Analog}\label{sample:0903}

One \hbox{radio-quiet} source in our sample, J0903+0708, shows notably different \xray\ properties from the other \hbox{radio-quiet} sources; 
it is much brighter in \hbox{X-rays} (see \S\ref{xray}). After the \chandra\ observations, we observed this source with the 
Hobby-Eberly Telescope (HET; Ramsey et~al. 1998) because it was a potential outlier. We obtained spectroscopy for J0903+0708 
using the G2 grating and the $1''$ slit of the \hbox{Low-Resolution} Spectrograph (LRS; Hill et~al. 1998) 
on the HET on 2010 November 5. Two HET spectra with spectral resolution $R\approx1000$ were extracted, 
each of which has an exposure time of 600 s. The spectra were calibrated to a standard
star (G191~B2B). Fig.~\ref{het_fig} shows the coadded HET spectrum 
and a comparison to the SDSS spectrum. The rest-frame wavelength range of the HET spectra is 1350$-$2250~\AA. The \hbox{rest-frame} 
UV emission-line measurements of the coadded HET spectrum are also listed in Table~\ref{qso_table}. The strengths of the emission  
lines (\ion{C}{4}, \ion{Si}{4}, \ion{C}{3}], and \ion{Fe}{3}) in the HET spectrum agree with those of the SDSS spectrum, 
while the blueshift of the \ion{C}{4} line in the HET spectrum is \hbox{$-610\pm300$~km~s$^{-1}$}, different at a $2.5\sigma$ 
level from the SDSS measurement ($-2300\pm600$~km~s$^{-1}$). 
In order to test for any possible variation of the \ion{C}{4} line, we performed a 
$\chi^2$ test on the \ion{C}{4} region (1486--1574~\AA) of the two spectra. These two spectra are consistent with each other
within the noise (\hbox{$\chi^2=132.3$} for 146 spectral bins). The HET spectrum has a much higher \hbox{signal-to-noise} 
ratio than the SDSS spectrum 
(see the inset of Fig.~\ref{het_fig} for the \ion{C}{4} line region), which indicates that the SDSS blueshift measurement 
was probably erroneous due to the SDSS spectrum's limited \hbox{signal-to-noise} ratio. With the improved 
HET data, J0903+0708 would not have passed our 
original selection criterion of \ion{C}{4} blueshift for \phl\ analogs (see \S\ref{sample:select}), 
and thus it should not be included in our sample. Note that J0903+0708 was one of the lowest \ion{C}{4} blueshift sources among our
candidate \phl\ analogs even with only the SDSS measurements. J0903+0708 is probably a regular WLQ (see \S\ref{discuss:wlq} 
for further discussions on the relation between \phl\ analogs and WLQs). We will therefore exclude this source from 
discussion hereafter unless noted. Nevertheless, we will report the \xray\ properties of J0903+0708 in \S\ref{xray}. Our final sample 
of \phl\ analogs includes 10 quasars; eight are radio quiet, while two are radio intermediate.

\subsection{Measurements of the UV Continua}\label{sample:cont}
\begin{figure*}[t]
    \centering
    \includegraphics[width=3.2in]{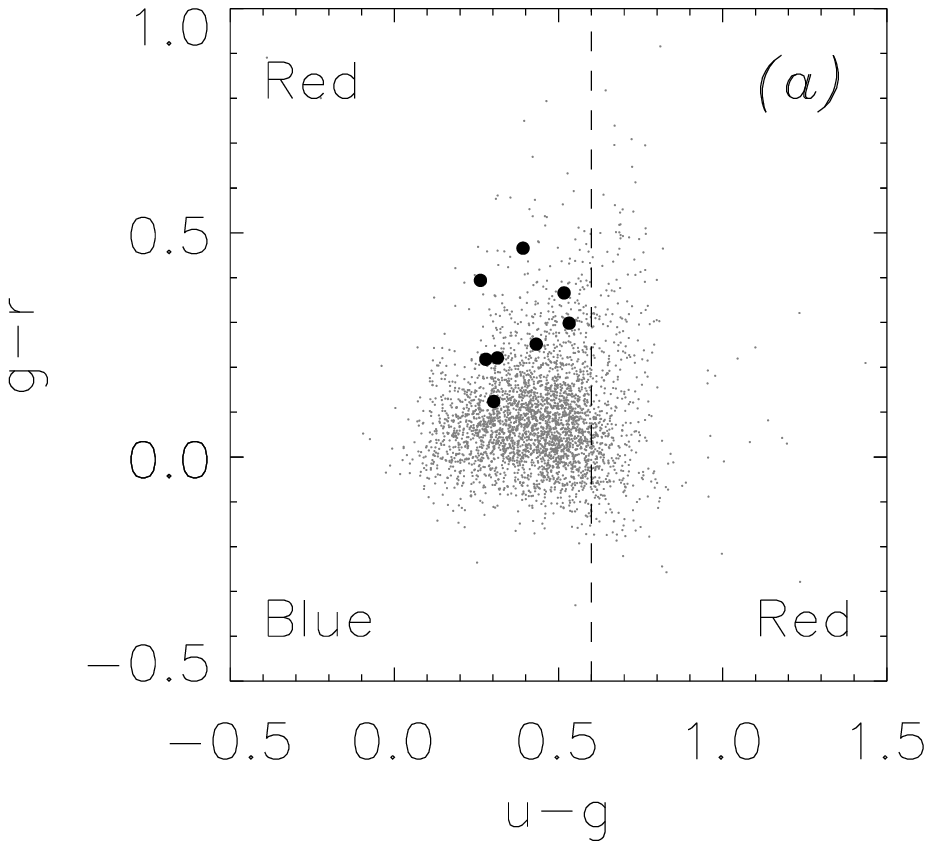}
    \includegraphics[width=3.2in]{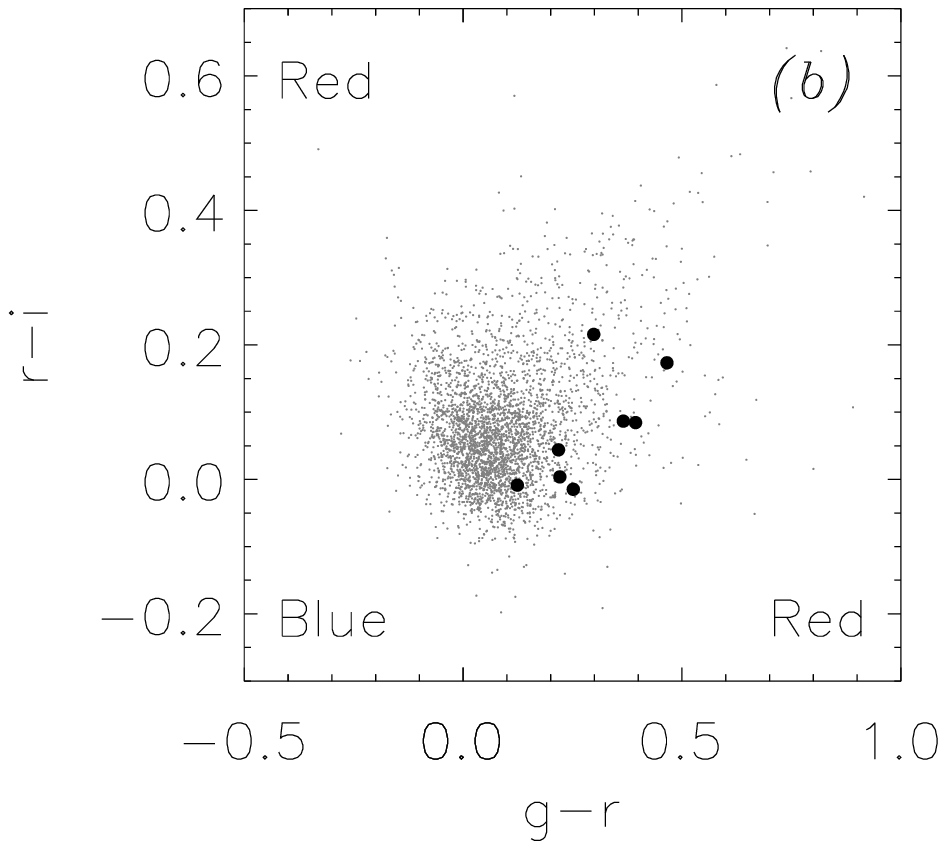}
    \includegraphics[width=3.2in]{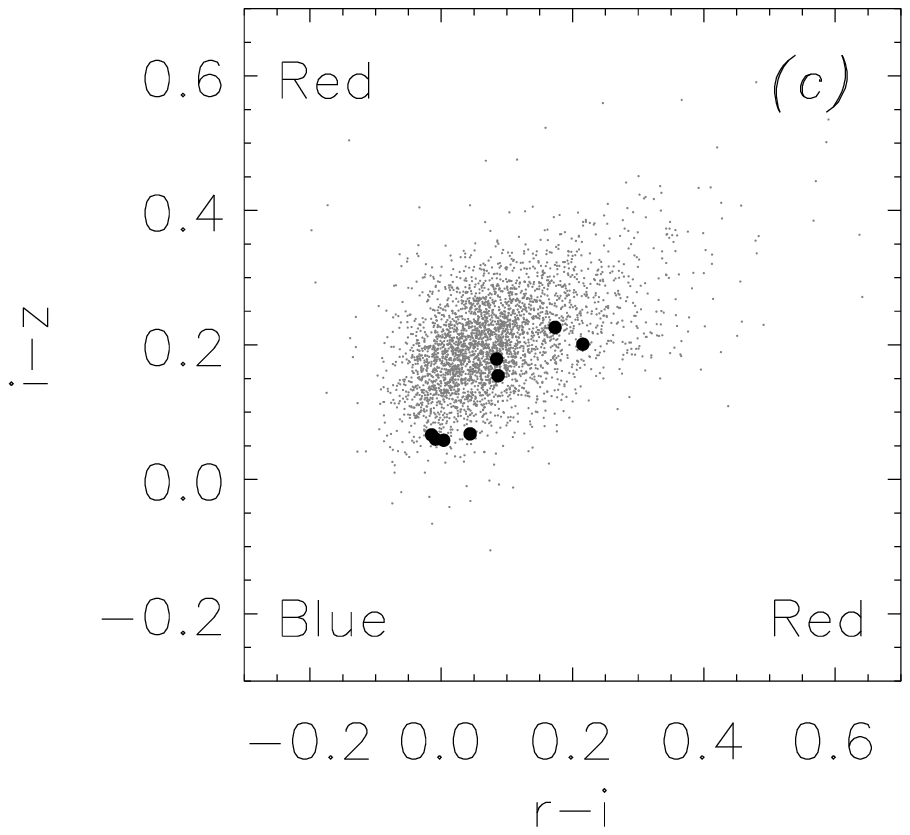}
    \caption{\footnotesize{The locations of our \phl\ analogs (black filled circles) in the SDSS color space: $(a)$ $u-g$ vs. $g-r$, 
             $(b)$ $g-r$ vs. $r-i$, and $(c)$ $r-i$ vs. $i-z$. The small grey dots show the SDSS color-selected quasars
             in the DR7 catalog (Schneider et~al. 2010) with similar redshifts ($2.125 < z < 2.385$) and luminosities ($M_i > -26.5$).
             The dashed line in panel $(a)$ shows an inclusion region $(u-g)<0.6$ of the SDSS color selection for quasar candidates.}
             \label{color_fig}}
\end{figure*}
We also estimated the spectral index of the presumed \pl\ continuum 
in the rest-frame 1200--2800~\AA\ range
($\alpha_\nu$, where \hbox{$f_\nu \propto \nu^{\alpha_\nu}$}) for each source by fitting the SDSS
spectra in four \hbox{line-free} regions using the method described in \S2.1 of Gibson et~al. (2008b). 
The results are listed in Table~\ref{qso_table}. 
The range of spectral indices for our candidate \phl\ analogs is
$-$0.97 to $-$0.28 with a mean value of $\langle\alpha_\nu\rangle=-0.61$.
This value is within the range seen for typical quasars, although it is
somewhat steeper (i.e., redder) than the ``canonical'' value of $\alpha_\nu=-0.5$ 
the optical and UV region because of the presence of the
``small blue bump'' due to the Balmer continuum and Fe emission 
(see discussion in \S2.1 of Strateva et~al. 2005;
also see Natali et~al. 1998, Schneider et~al. 2001, Vanden Berk et~al. 2001).

The somewhat redder than average nature of our quasars is evidenced by
their $\Delta (g-i)$ values (the $g-i$ color minus the average $g-i$ color
for quasars at the same redshift) taken from Schneider et~al. (2010).
On average, our sample is redder than $\approx75\%$ of other quasars at the same
redshift, with $\langle\Delta(g-i)\rangle=0.28$.  About a quarter of this excess can
be attributed to weak \ion{C}{4} emission in the $g$ band and about a
quarter to strong \ion{Fe}{2} emission in the $i$ band (see Fig.~\ref{spec_fig}).  The remaining
half corresponds to the redder continuum expected for quasars with
spectral index $\alpha_\nu=-0.61$ instead of $\alpha_\nu=-0.5$.

In spite of their somewhat redder colors, we do not expect substantial incompleteness in the SDSS 
selection of \phl\ analogs. Our \phl\ analogs lie within the loci of $u-g$, $g-r$, $r-i$, 
and $i-z$ colors for SDSS \hbox{color-selected} quasars at similar redshifts and luminosities, but 
they are somewhat offset from the regions with the highest density of quasars 
(see Fig.~\ref{color_fig}). They are within the inclusion region of $(u-g)<0.6$ for SDSS color 
selection of quasar candidates (see \S3.5.2 of Richards et~al. 2002). Our objects are also in a 
redshift range where the SDSS color selection algorithm has high completeness ($\approx75\%$; see 
Fig.~6 of Richards et~al. 2006).


\section{X-ray Data Analysis}\label{xray}

The eight \chandra\ Cycle~11 targets were observed with the S3 CCD of the Advanced
CCD Imaging Spectrometer (ACIS; Garmire et~al. 2003). Data reduction was performed
using the standard CIAO v4.2 procedures. \xray\ images were generated for 
the observed-frame soft (\hbox{$0.5$--$2.0$}~keV), hard (\hbox{$2.0$--$8.0$}~keV), 
and full (\hbox{$0.5$--$8.0$}~keV) 
bands using {\it ASCA} grade 0, 2, 3, 4, and 6 events. Source detections were
performed with the {\sc wavdetect} algorithm (Freeman et~al. 2002) using a 
detection threshold of $10^{-5}$ and wavelet scales of $1$, $\sqrt{2}$, $2$, 
$2\sqrt{2}$, and $4$ pixels. All targets, except J0848+5408 and J1230+3320, were detected in
at least one band within $0.5''$ of the object's optical coordinates. 
Aperture photometry was performed using the IDL {\sc aper} routine with an aperture radius of 
$1.5''$ for the source ($\approx95\%$ enclosed energy for soft band, $\approx90\%$ enclosed 
energy for hard band), and inner and outer annulus radii 
of $3.0''$ and $4.5''$ for background
subtraction, respectively. All background regions are free of \xray\ sources. For 
undetected sources, the upper limits
upon \xray\ counts were determined using the method of Kraft et~al.~(1991) for the low-count 
scenario of $N<10$, where $N$ is the total counts within the source aperture. Aperture 
corrections were applied to the \xray\ counts or to their upper limits.
The \xray\ counts in the three bands defined above, as well as the band ratio (defined as
the hard counts divided by the soft counts) and effective \pl\ photon index, for all newly
observed and archival sources,
are reported in Table~\ref{cts_table}. The effective \pl\ photon index was calculated 
from the band ratio using
the \chandra\ PIMMS\footnote{http://cxc.harvard.edu/toolkit/pimms.jsp} tool, assuming a 
\pl\ model with the Galactic absorption. 

One archival source, J1521+5202, was observed by \chandra\ in Cycle 7; these results
were reported in Just et~al.~(2007). We reanalyzed these \chandra\ data and found 
consistent results with those in Just et~al.~(2007). This extremely optically 
luminous ($M_i=-30.19$, see Fig.~\ref{zMi_fig}) source is exceptionally \xray\ weak;
it is not detected in 
the soft band. The other \chandra\ archival source, J1219+1244, was serendipitously
covered in Cycle 8, with an off-axis angle of $3.0'$. 
It is undetected in all three bands. The aperture for this source ($2.1''$ radius) 
was taken to be 
the 95\% enclosed-energy radius at $1.497$~keV according to the point spread function 
(PSF) of the \hbox{ACIS-S} detector. 

The \xmm\ archival source, J1230+2049, was serendipitously observed on 2001~July~1. 
The data were processed using standard
\xmm\ Science Analysis System (v10.0.0) routines. Only data from the \verb+MOS+ detectors were
used because this source is located on a CCD edge of the \verb+pn+ detector. The event files were 
filtered by removing periods of background flaring in which the count rate exceeded
$0.35$~s$^{-1}$ for events with energies above 10~keV. Only 3\% of the exposure time was removed via 
this filtering procedure. This source is undetected in all three
bands using the {\sc eboxdetect} procedure. The aperture for photometry ($9.3''$ radius) was taken to be 
the 50\% enclosed-energy radius (to avoid a nearby \xray\ source) at $1.5$~keV according to the PSF of 
the \verb+MOS+ detectors at an off-axis angle of $6.0'$. The upper limits upon \xray\ counts were also
determined using the Kraft et~al. (1991) method. 

Table~\ref{aox_table} lists the key \xray, optical, and radio properties of our sample:

\noindent Column (1): The SDSS~J2000 equatorial coordinates for the quasar.



\noindent Column (2): The apparent $i$-band magnitude of the quasar using the 
SDSS quasar catalog BEST photometry, $m_{i}$.

\noindent Column (3): The absolute $i$-band magnitude for the quasar, $M_{i}$, from the 
SDSS DR7 quasar catalog (Schneider et~al. 2010), which was calculated by correcting for 
Galactic extinction and assuming a \pl\ spectral index of $\alpha_\nu\;=\;-0.5$.

\noindent Column (4): The Galactic neutral hydrogen column density obtained 
with the \chandra\ COLDEN\footnote{http://cxc.harvard.edu/toolkit/colden.jsp} 
tool (Dickey \& Lockman 1990; Stark et~al. 1992), in units of $10^{20}$~cm$^{-2}$.

\noindent Column (5): The count rate in the observed-frame soft \xray\ band 
($0.5$--$2.0$~keV), in units of $10^{-3}$~s$^{-1}$. For the two off-axis sources (J1219+1244 and
J1230+2049), the upper limits upon count rate are corrected for vignetting using 
exposure maps. 

\noindent Column (6): The Galactic absorption-corrected flux in the 
observed-frame soft \xray\ band (\hbox{$0.5$--$2.0$~keV}) obtained with
\chandra\ PIMMS tool, in units of $10^{-14}$~erg~cm$^{-2}$~s$^{-1}$. An 
absorbed \pl\ model was used with a photon index $\Gamma=2$, which is typical 
for quasars, and the Galactic neutral hydrogen column density 
($N_H$, given in Column~4).

\noindent Column (7): The Galactic absorption-corrected flux density at 
rest-frame 2 keV obtained with PIMMS, in units of $10^{-32}$ erg cm$^{-2}$ s$^{-1}$ Hz$^{-1}$.

\noindent Column (8): The logarithm of the Galactic absorption-corrected quasar 
luminosity in the rest-frame $2-10$~keV band.

\noindent Column (9): The continuum flux density at rest-frame 2500~\AA~ in 
units of $10^{-27}$ erg~cm$^{-2}$~s$^{-1}$~Hz$^{-1}$, which was obtained by fitting the 
SDSS continuum in line-free regions using the method described in \S2.1 of Gibson et~al. (2008b).
The SDSS spectra were corrected for fiber light loss and Galactic extinction.

\noindent Column (10): The logarithm of the 
monochromatic luminosity at rest-frame 2500~\AA, calculated from the flux density at 
rest-frame 2500~\AA. 

\noindent Column (11): The X-ray-to-optical \pl\ slope, given by
\begin{equation}
    \alpha_{\rm ox} = \frac{{\rm log}(f_{\rm 2\;keV} / 
    f_{2500\mbox{\rm~\scriptsize\AA}})}{{\rm log}(\nu_{\rm 2\;keV} / \nu_{2500\mbox{\rm~\scriptsize\AA}})}
    = 0.384\ {\rm log} \bigg(\frac{f_{\rm 2\;keV}}{f_{2500\mbox{\rm~\scriptsize\AA}}}\bigg) .
\end{equation}
All measures of flux density used are per unit frequency. Note our UV and \xray\ measurements
were not simultaneous. 

\noindent Column (12): \daox, defined as
\begin{equation}
    \Delta\alpha_{\rm ox} = \alpha_{\rm ox(measured)} - \alpha_{\rm ox(expected)}.
\end{equation}
The expected \aox\ value for a typical quasar is calculated 
from the $\alpha_{\rm ox}$-$L_{2500\mbox{\rm~\scriptsize\AA}}$ correlation given as 
Equation (3) of Just et~al.~(2007). The statistical significance of this 
difference, given in parentheses, is in units of $\sigma$, which is 
given in Table~5 of Steffen et~al.~(2006) as the RMS for \aox\ for 
several ranges of luminosity. Here, $\sigma$~=~0.146 for 
31~$<$~log$L_{2500\mbox{\rm~\scriptsize\AA}}$~$<$~32 and $\sigma$~=~0.131 for 
32~$<$~log$L_{2500\mbox{\rm~\scriptsize\AA}}$~$<$~33.

\noindent Column (13): The factor of \xray\ weakness, converted from the \daox\ values in Column~(12), 
quantifying the \xray\ weakness of our targets compared to a typical quasar with similar UV/optical
luminosity, which is calculated as $f_{\rm x-weak}=10^{-\Delta\alpha_{\rm ox}/0.384}\approx403^{-\Delta\alpha_{\rm ox}}$. 
A source with a \daox\ value of $-0.384$ has an \xray\ flux only $\approx10\%$ that of typical quasars, 
corresponding to an \xray\ weakness factor of $\approx10$. 

\noindent Column (14): The radio-loudness parameter, given by 
\begin{equation}
    R = \frac{f_{\rm 5\;GHz}}{f_{4400\mbox{\rm~\scriptsize\AA}}}{\rm .}
\end{equation}
The denominator, $f_{4400\mbox{\rm~\scriptsize\AA}}$, was found via extrapolation from 
$f_{2500\mbox{\rm~\scriptsize\AA}}$ using a UV/optical \pl\ slope of $\alpha_\nu$ = $-0.5$. 
We also utilized the individual measurements of $\alpha_\nu$ for each source given in \S\ref{sample:cont}, and we found
that these would not change the source classifications as radio quiet ($R<10$) or radio intermediate
($10<R<100$).
The numerator, $f_{5\;{\rm GHz}}$, was found using a radio \pl\ slope of 
$\alpha_\nu$ = $-0.8$ and a flux at 20 cm, $f_{20\;{\rm cm}}$. For sources detected by the 
{\it FIRST} survey (Faint Images of the Radio Sky at Twenty-Centimeters; Becker et~al. 1995), 
$f_{20\;{\rm cm}}$ was taken from the {\it FIRST} source
catalog. For sources covered but not detected by the {\it FIRST} survey, the upper limits for 
 $f_{20\;{\rm cm}}$ were calculated as three times of the RMS noise in a $0.5'\times0.5'$ 
{\it FIRST} image cutout at the object's coordinates. J0000+1356, not located in the 
{\it FIRST} survey area, is covered but not detected by the {\it NVSS} (The NRAO VLA Sky Survey; 
Condon et~al. 1998).
The upper limit for $f_{20\;{\rm cm}}$ for this source is taken as 1.35~mJy, which is three times 
the RMS noise of the {\it NVSS}. Two sources in our sample, J1454+0324 and J1618+0704, 
are \hbox{radio-intermediate} quasars; their \xray\ emission may contain 
significant contributions from associated jets. 


\section{Results and Discussion}\label{discuss}

\subsection{Relative X-ray Brightness}\label{discuss:daox}
\begin{figure}[t]
    \centering
    \includegraphics[width=3.5in]{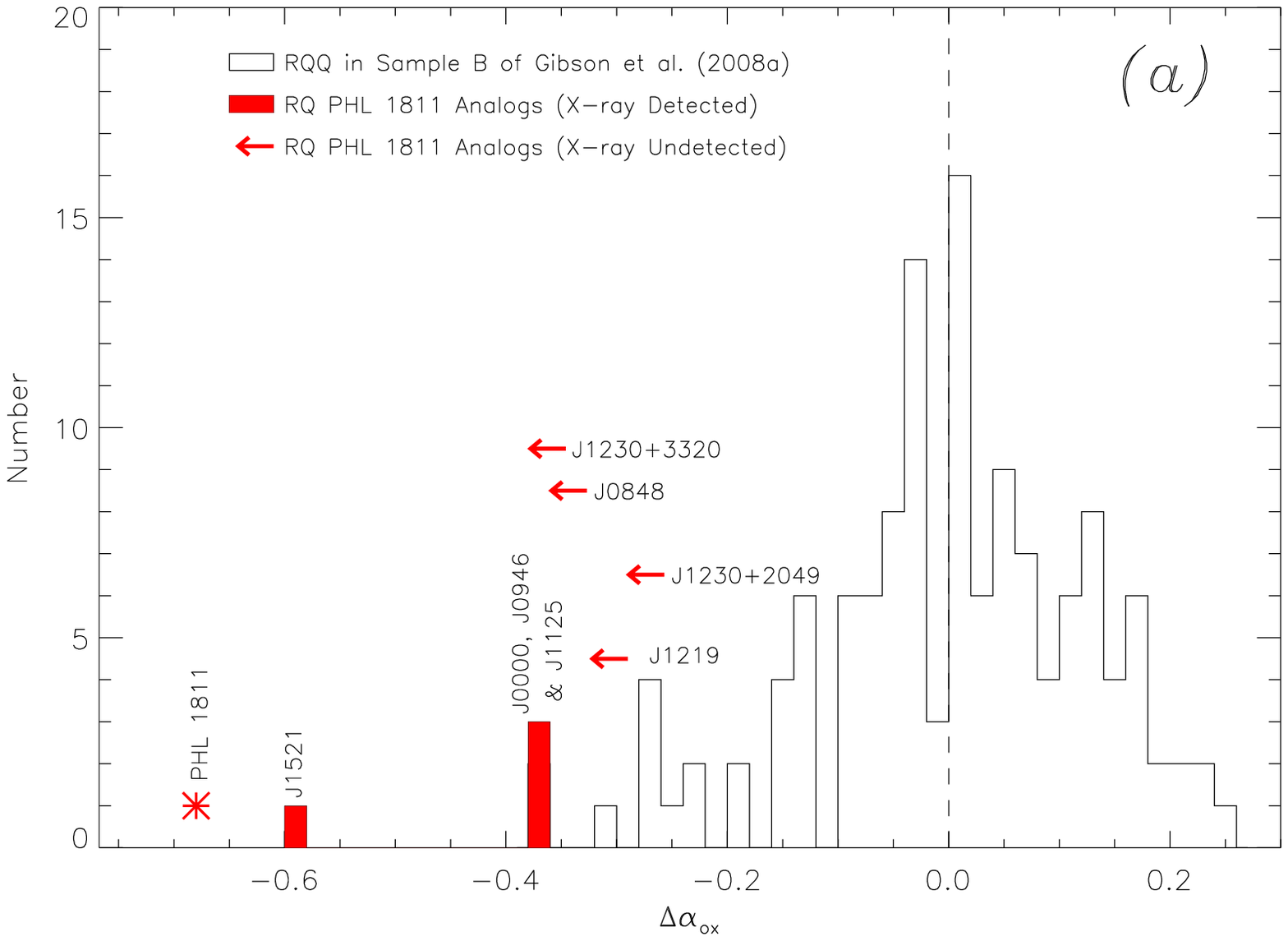}
    \includegraphics[width=3.5in]{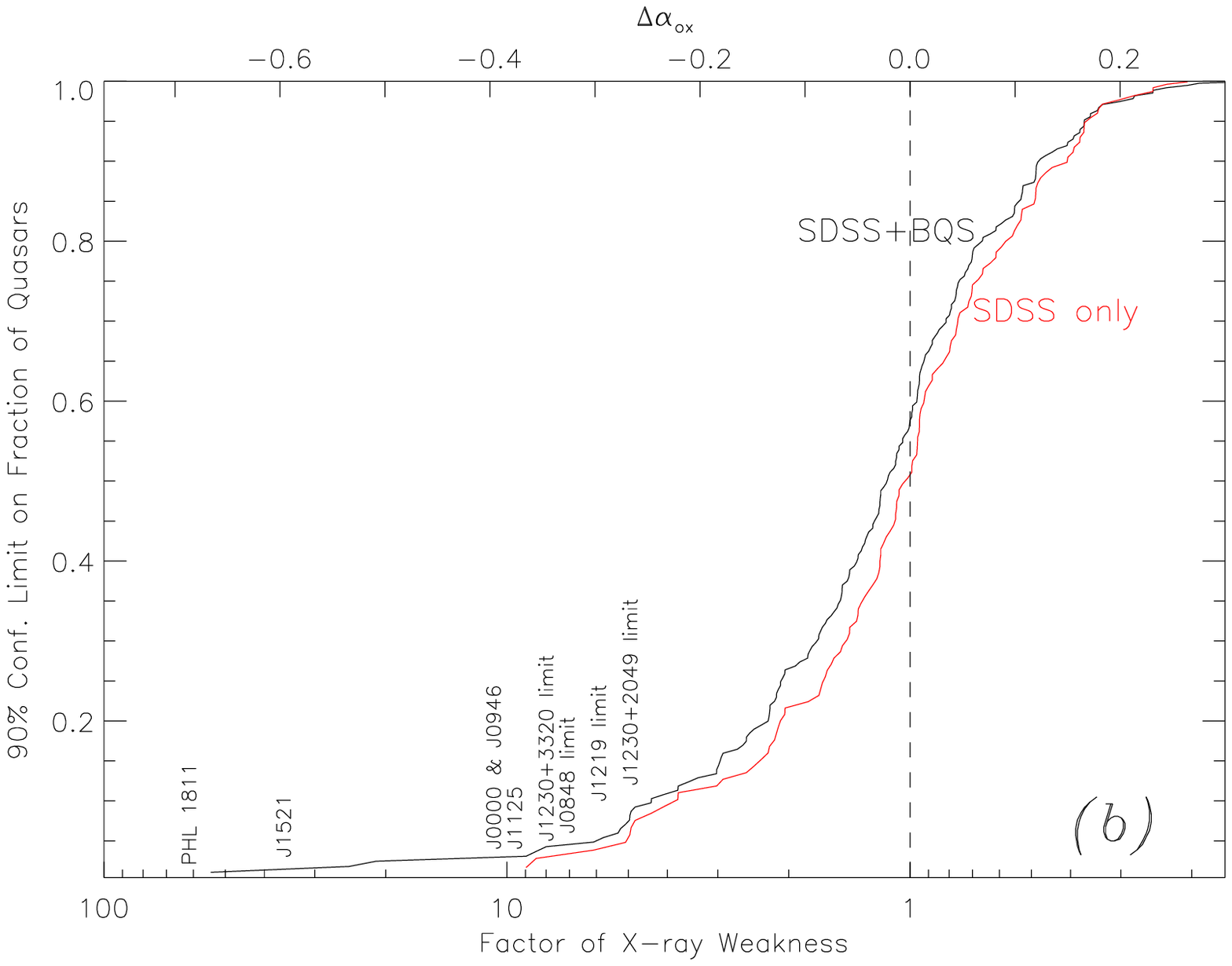}   
    \caption{\footnotesize{$(a)$: Distribution of \daox\ values for \hbox{radio-quiet} objects ($R<10$) in our sample of \phl\ analogs,
             compared to that of the 132 \hbox{radio-quiet}, \hbox{non-BAL} quasars in Sample~B of Gibson et al. (2008a). The red shaded 
             histogram and leftward arrows represent \hbox{radio-quiet} sources in our sample which are detected and undetected 
             in \hbox{X-rays}, respectively. The unshaded histogram shows the 
             \hbox{radio-quiet}, \hbox{non-BAL} quasars in Sample~B of Gibson et al. (2008a). $(b)$: 90\% confidence upper limit on the 
             fraction of SDSS (red curve) and SDSS+BQS (black curve) \hbox{radio-quiet}, \hbox{non-BAL} quasars that are \xray\ weak by a 
             given factor (adapted from Fig.~5 of Gibson et al. 2008a). The factors of \xray\ weakness for \phl\ analogs 
             in our sample are shown. The \daox\ values and the corresponding \xray\ weakness factors 
             in the two panels are aligned with each other. Source names for \phl\ analogs are labelled in both panels in 
             the format of 'J$hhmm$' for brevity, except for J1230+3320 and J1230+2049. The dashed vertical line in each panel 
             shows \daox $=0$, corresponding to an \xray\ weakness factor of unity.}
             \label{daox_fig}}
\end{figure}

The \daox\ parameter (see Column~12 of Table~\ref{aox_table}) quantifies the relative \xray\ brightness of a quasar with respect
to a typical \hbox{radio-quiet} quasar with the same UV luminosity. Fig.~\ref{spec_fig} shows the SDSS
spectra of our \phl\ analogs (and also J0903+0708; see \S\ref{sample:0903}) ordered by \daox. No strong trends are apparent among the 
\phl\ analogs between \daox\ and the UV \hbox{emission-line} characteristics (see \S\ref{discuss:corr} for quantitative 
correlation analyses). Two \hbox{radio-intermediate} sources in our sample\footnote{``Our sample'' here refers to 
our 10 \phl\ analogs.}  are \xray\ bright (see further discussion in \S\ref{discuss:radio}).
The distribution of \daox\ values for the eight \hbox{radio-quiet}
\phl\ analogs is shown in Fig.~\ref{daox_fig}$(a)$, compared to that for the 132
\hbox{radio-quiet}, \hbox{non-BAL} quasars in Sample~B of Gibson et~al. (2008a).\footnote{Sample~B of Gibson et al. (2008a) includes 139 objects characterized as radio-quiet, non-BAL quasars.  
However, we have identified a further seven of these sources as BAL quasars (with C~{\sc iv}~${\rm BI}_0 > 0$) 
in the SDSS DR5 BAL quasar catalog (Gibson et al. 2009a; the sources are J0050$-$0057, J0756+4458, J1212+1520, J1426+3517, J1426+3753, J1430+3221, and J1434+3340). For four of these seven sources, 
the classification depends on the reconstruction of the C~{\sc iv} emission-line profile 
(see the discussion of the {\it Emlost} flag in \S2 of Gibson et al. 2009a). In the other three cases, BAL classification is
sensitive to the continuum model placement.  The continuum models in Gibson et al. (2009a) were generally different from 
those in Gibson et al. (2008a), so the identifications of BALs may differ in these cases. An additional six sources 
(J0855+3709, J1049+5858, J1245$-$0021, J1338+2922, J1427+3241, and J1618+3456) appear to have possible BAL or \hbox{mini-BAL} 
features in our visual inspections of their SDSS spectra. The seven quantitatively identified BAL quasars are removed 
in the following analysis. The six visually identified BAL or mini-BAL quasars will be kept in Sample B, but shown by 
different symbols in subsequent figures.} The Sample~B quasars were all
serendipitously detected by \chandra\ (with off-axis angles greater than $1'$). They are representative of typical
SDSS quasars in the redshift range of $1.7\leq z \leq 2.7$. We use a Peto-Prentice test 
(e.g., Latta 1981), implemented in the Astronomy Survival Analysis (ASURV) package (e.g. Lavalley et~al. 1992), 
to assess whether these two samples follow the same distribution. The Peto-Prentice test
is our preferred approach because it is the least affected by the factors of unequal sample sizes or
different censoring patterns which exist in our case (e.g., Latta 1981). The distribution of 
\daox\ values for our sample is found to be significantly different from that for 
typical SDSS quasars; the probability of the null hypothesis (the two samples following the same distribution)
is reported to be $1\times10^{-28}$ by the Peto-Prentice test. Other two-sample tests for censored data 
(e.g., Gehan, logrank, and Peto-Peto)
give similar results (see Table~\ref{twost_table}). The two-sample test results do not change materially 
if J0903+0708 (see \S\ref{sample:0903}) is included (also see Table~\ref{twost_table}).

The mean value of \daox\ for \hbox{radio-quiet} \phl\ analogs in our sample is $-0.423\pm0.049$, 
which is calculated using the Kaplan-Meier estimator\footnote{The Kaplan-Meier estimator is applicable to 
censored data and reliable for small-sample cases (E.~D.~Feigelson 2010, private communication).} 
(also implemented in the ASURV package), while the mean \daox\ value for \hbox{radio-quiet}, \hbox{non-BAL} quasars in 
Sample~B of Gibson et~al. (2008a) is $-0.001\pm0.011$. Therefore, \hbox{radio-quiet} \phl\ analogs in our sample are 
anomalously \xray\ weak by a mean factor of $\approx13$. If J0903+0708 is included, the 
mean value of \daox\ changes to $-0.384\pm0.057$, corresponding to a mean \xray\ weakness factor of $\approx10$.
All of our \hbox{radio-quiet} \phl\ analogs are \xray\ weak by factors $>4.8$; we only have 
lower limits on the range of \xray\ weakness factors because four of the \phl\ analogs are undetected in \hbox{X-rays}. 
Our \xray\ detected sources show that the \xray\ weakness factors span at least the range from 8.7 to 34.5.
\subsection{Average \xray\ Properties via Stacking Analyses}\label{discuss:stack}

For the three \hbox{radio-quiet} \phl\ analogs not detected by \chandra\ (J0848+5408, J1219+1244, and J1230+3320), we performed 
a stacking analysis to obtain an average 
constraint on their relative \xray\ weakness. We added the total counts and background counts of the three sources in 
their apertures for the full band, soft band, and hard band. For each band, we then calculated the Poisson 
probability of the total counts arising from a fluctuation of the corresponding background counts. If
this probability is less than $0.1\%$, we take the stacked source to be detected in that band. The stacked source is detected in 
the full band, with $4.42$ total counts and $0.35$ background counts; the Poisson probability is $1\times 10^{-4}$. 
The stacked source is not detected in the soft or hard band, so a band-ratio analysis is not
possible. The combined 
effective exposure time for the stacked source is 23.3 ks. After an aperture correction was applied, the average full-band
count rate of the stacked source is $1.84\times 10^{-4}$ counts~s$^{-1}$. We adopted the average values
of the parameters for the three sources, such as Galactic $N_H$, $z$, and $f_{2500~\mathring{\rm{A}}}$, to calculate
the \xray\ properties of the stacked source. Under the assumption of a Galactic-absorbed \pl\ model with $\Gamma=2$, 
the average \xray\ flux density at rest-frame 2~keV is $2.90\times 10^{-33}$~erg~cm$^{-2}$~s$^{-1}$~Hz$^{-1}$, while the 
average rest-frame \xray\
luminosity from 2--10~keV is $2.75\times 10^{43}$~erg~s$^{-1}$. The average \aox\ value for the stacked source
is $-2.26$. The \daox\ value is $-0.57$, corresponding to an \xray\ weakness factor of $30.5$ (cf. Fig.~\ref{daox_fig}). 
At least on average, 
these undetected \phl\ analogs appear to be only a few times \xray\ weaker than those we have detected individually. 
\phl\ itself has similar \xray\ weakness ($\approx30$--$100$) to these undetected \phl\ analogs on average (as does
J1521+5202). 

We also performed a stacking analysis of the four \hbox{radio-quiet} \phl\ analogs that are detected by \chandra\ 
(J0000+1356, J0946+2744, J1125+5028, and J1521+5202) to constrain their average \xray\ spectral properties. 
We added the counts 
for each source in soft band and hard band. The total \hbox{background-subtracted} counts are $12.9^{+4.7}_{-3.5}$ in the soft band 
and $7.7^{+3.9}_{-2.7}$ in the hard band. The band ratio is $0.60^{+0.38}_{-0.26}$, corresponding to an effective photon index of 
$\Gamma=1.22^{+0.52}_{-0.45}$ using the average Galactic neutral hydrogen column density of these four sources. 
This photon-index value suggests a harder average \xray\ spectrum for the \xray\ detected, 
\hbox{radio-quiet} \phl\ analogs than that for typical \hbox{radio-quiet} quasars. This 
average \xray\ spectrum is also harder than the \xray\ spectrum of \phl\ ($\Gamma\approx2.2$; Leighly et~al. 2007a). However, even
the quoted $68\%$ confidence error bars are large due to limited counts. If quoting $90\%$ confidence error bars, the 
photon index is $\Gamma=1.22^{+0.85}_{-0.66}$. If we furthermore add the counts from the three \phl\ analogs undetected 
by \chandra\ (see the previous paragraph), the average effective photon index becomes $\Gamma=1.10^{+0.45}_{-0.40}$
(68\% confidence level), or $1.10^{+0.77}_{-0.59}$ (90\% confidence level). 

\subsection{Demographic Constraints}\label{discuss:demo} 

Our sample selection and observational results allow us to investigate the demographics of \phl\ analogs in the 
total quasar population. Our \phl\ analogs were selected from \hbox{high-redshift} ($2.125\leq z\leq2.385$), optically 
bright ($m_r\leq18.8$) objects classified as \verb+QSO+ or \verb+HIZ_QSO+ in the SDSS DR7 CAS 
(see \S\ref{sample:select}). There are 1358 \hbox{radio-quiet}, 
\hbox{non-BAL} SDSS DR7 quasars within the above redshift and magnitude ranges. Our sample of \phl\ analogs contains eight 
\hbox{radio-quiet} and two \hbox{radio-intermediate} sources. The \hbox{radio-quiet} \phl\ analogs thus appear to be only 
$8/1358=(0.6^{+0.5}_{-0.3})\%$ (at the 90\% confidence level) of the total \hbox{radio-quiet}, \hbox{non-BAL} quasar population. 
Considering the potential incompleteness of our sample selection, we take 0.3\% as a lower limit on the fraction 
of \phl\ analogs in the \hbox{radio-quiet}, \hbox{non-BAL} quasar population. On the other hand, 
Fig.~\ref{daox_fig}$(b)$ shows upper limits on the fraction of optically selected, \hbox{radio-quiet}, \hbox{non-BAL} 
quasars compiled from samples using the SDSS only and the SDSS+BQS (Bright Quasar Survey; Schmidt \& Green 1983) 
that are \xray\ weak by a given factor (adapted from Fig.~5 of Gibson et~al. 2008a). Only $\leq2.8\%$ (at the 90\% 
confidence level) of those quasars are as \xray\ weak (\daox$=-0.427$; see \S\ref{discuss:daox}) as the \hbox{radio-quiet} 
sources in our sample. In summary, the fraction of \hbox{radio-quiet} \phl\ analogs should be within the range of $0.3\%$--$2.8\%$. 
The upper limit of $2.8\%$ is obtained solely based on the factor of \xray\ weakness. 
Since our selection of \phl\ analogs should not have strong incompleteness (a factor of $\lesssim2$), 
we estimate the fraction of \phl\ analogs in the total quasar population to be $\lesssim1.2\%$. 

Radio-quiet \phl\ analogs, which are all \xray\ weak in our sample, only compose a small fraction of the total 
quasar population. They do not appear to present fractionally significant difficulties to the utility of \xray\ surveys 
for finding AGNs throughout the Universe (see \S\ref{intro}). The majority ($\gtrsim80\%$) of BAL quasars are also \xray\ weak 
owing to intrinsic \xray\ absorption (e.g., Gibson et~al. 2009a). Considering the fraction of BAL quasars 
($\approx15\%$--$20\%$; e.g., Gibson et~al. 2009a) found in SDSS quasar samples, \xray\ weak BAL quasars outnumber 
\xray\ weak \phl\ analogs by a factor of $\gtrsim10$.

\subsection{Intrinsic \xray\ Weakness vs. \xray\ Absorption}\label{discuss:intrinsic}

Our \hbox{radio-quiet} \phl\ analogs could be \xray\ weak either because they are 
intrinsically \xray\ weak, perhaps due to quenching of the ADC (see \S\ref{intro}), 
or because they are \xray\ absorbed. In this section, we will discuss the 
strengths and weaknesses of each of these hypotheses. 

There are three indirect arguments supporting the intrinsic \xray\ weakness scenario. 
First, as discussed in \S\ref{sample:select}, we have selected against objects with detectable broad \ion{C}{4} absorption
(or any other broad absorption lines, including \ion{Si}{4} and \ion{Mg}{2}), since 
most heavily \xray\ absorbed type~1 quasars show broad \ion{C}{4} absorption 
(e.g., Brandt et~al. 2000; Gibson et~al. 2008a; Page et~al. 2010). If we assume the
\phl\ analogs have underlying \xray\ spectra with $\Gamma=2$, the level of intrinsic neutral absorption would need to be
\hbox{$N_H \approx 5\times 10^{23}$~cm$^{-2}$} ($\gtrsim 2\times 10^{23}$~cm$^{-2}$ to 
$\gtrsim 8\times 10^{23}$~cm$^{-2}$) at $z\approx2.2$ to generate the observed \xray\ weakness 
factor of 13 (and the range of $>4.76$ to $\geq 34.5$). Three of our eight \hbox{radio-quiet} \phl\ analogs 
(J0848+5408, J1219+1244, and J1230+2049) have narrow associated \ion{C}{4} absorption lines
(within $\pm$5000~km~s$^{-1}$ of the quasar redshift).
Quasars with associated NALs generally have similar \xray\ properties to unabsorbed quasars and are not heavily \xray\ absorbed 
(e.g., Misawa et~al. 2008; Chartas et~al. 2009). The intrinsic \xray\ absorption for quasars with associated NAL systems 
is typically $N_H \lesssim 10^{22}$~cm$^{-2}$, much lower than the level required for our \hbox{radio-quiet} \phl\ analogs. 
The frequency of associated NAL systems in our sample ($3/8=38\%$) is 
consistent with that of typical quasars (see \S3.2.1 of Ganguly \& Brotherton 2008). 
For the total number of NAL systems (both associated and intervening), the work of Nestor et al. (2008) predicts 
that between Ly$\alpha$ and \ion{C}{4} in the spectra of our eight \hbox{radio-quiet} \phl\ analogs we should find 16 \ion{C}{4}
absorbers with FWHM $<$ 600~km~s$^{-1}$ and $W_r(\lambda 1548) \leq -0.3$\,\AA. In fact, we find 10. Within the limitations 
of \hbox{small-number} statistics, there is nothing unusual about the total number of NAL systems in our \phl\ analogs.
One source, J1219+1244, has a \ion{C}{4} absorber with a total velocity width of 
$1100$~km~s$^{-1}$,\ just above the $1000$~km~s$^{-1}$ threshold to be termed a \hbox{mini-BAL}
quasar. 
Mini-BAL quasars are also generally not 
strongly \xray\ absorbed (e.g., Gibson et~al. 2009b; Wu et~al. 2010a). The intrinsic \xray\ absorption
of \hbox{mini-BAL} quasars is also typically at the level of $N_H \lesssim 10^{22}$~cm$^{-2}$ (Wu et~al. 2010a).
While it is perhaps possible that a type~1 quasar could be strongly \xray\ absorbed without showing
notable UV absorption lines (see below), 
the available empirical evidence indicates that our removal of quasars with BALs 
should eliminate the vast majority of strongly \xray\ absorbed quasars.  

Second, the quasars in our sample show the blue UV/optical continua typical
of type~1 quasar spectra.  Our sample was chosen to avoid objects with
detectable spectral curvature indicative of strong dust reddening.
Nonetheless, the somewhat redder than average colors of our sample (see \S\ref{sample:cont})
suggest that we cannot rule out an average dust reddening of up to
$E(B-V) \approx 0.05$ magnitudes.  For an SMC dust-to-gas ratio (Bouchet et~al. 1985), such
reddening would imply a column density of \hbox{$N_H \approx 2.2 \times 10^{21}$ cm$^{-2}$}.
Again, this column density is $\gtrsim 100$ times too small to explain the \xray\ weakness of
our \hbox{radio-quiet} \phl\ analogs. Note, however, that this is not a strong argument against \xray\ absorption, 
since many BAL quasars (which usually have heavy \xray\ absorption) also have blue UV/optical continua
(e.g., Trump et~al. 2006). 

Third, the unusual UV emission-line properties of \phl\ can be explained naturally by an intrinsically 
\xray\ weak SED (Leighly et~al. 2007b). Such a continuum lacks adequate photons to create highly ionized ions,
which results in weak high-ionization lines (e.g., \ion{C}{4}, \ion{Si}{4}). The temperature of the gas is 
too low to excite semi-forbidden lines (e.g., \ion{C}{3}]). The strong \ion{Fe}{2} and \ion{Fe}{3} 
emission can also be qualitatively explained by the \phl-like SED (Leighly et~al. 2007b). 
Since our \phl\ analogs have the same unusual UV emission-line 
properties, Occam's razor would suggest their SEDs are also probably intrinsically \xray\ weak.



In spite of the indirect arguments above, it is notable that the average \xray\ spectrum for our radio-quiet \phl\ analogs 
appears to be harder than for typical radio-quiet quasars and for \phl\ itself (see \S\ref{discuss:stack}), which hints 
at the presence of heavy \xray\ absorption.\footnote{The harder average spectrum for these \xray\ weak objects could
also perhaps be explained if their \xray\ spectra are reflection-dominated, similar to what may apply for PG~0844+349 in an \xray\ weak 
state (Gallo et al. 2011).} It is possible that \xray\ absorption could also lead to the \xray\ 
weak SEDs required to create the unusual UV emission lines of our \phl\ analogs. However, if heavy \xray\ absorption is 
responsible, it must lie closer to the SMBH than the BELR does, so that this \hbox{emission-line} region ``sees'' a soft ionizing 
SED and produces a \phl-like UV emission-line spectrum. It must also 
have a remarkable lack of accompanying \ion{C}{4}, \ion{Si}{4}, or \ion{Mg}{2} BALs (or even \hbox{mini-BALs} in all but one case). 
These requirements might be achieved by a line of sight that passes through ``shielding gas'' but not a typical 
UV absorbing wind (e.g., Gallagher et al. 2005). The shielding gas, which resides interior to the UV absorbing wind, 
is sufficiently highly ionized that it is transparent 
to UV photons but absorbs soft \xray\ photons, preventing overionization of the wind 
(e.g., Murray et al. 1995). To explain \phl\ analogs, 
the shielding gas could have an atypically larger covering factor of the \xray\ continuum source that prevents most of the 
\hbox{X-rays} from reaching the BELR; we return to this idea in 
\S\ref{discuss:wlq}. The shielding gas might betray its presence via very highly ionized absorption, 
either in the UV or in \hbox{X-rays}.  For example, Telfer et al. (1998) report that the \xray\ weak quasar SBS~1542+541 has BAL 
troughs much stronger in \ion{O}{6} and \ion{Ne}{8} than in \ion{N}{5} or \ion{C}{4} (see their Fig.~5), and \ion{O}{6} 
and \ion{Ne}{8} have been seen in narrower absorption systems as well (e.g., Scott et al. 2004).

In fact, \phl\ itself has a narrow \ion{O}{6} absorber at $z=0.192186$ (consistent with the \hbox{Balmer-line} redshift of \phl\ 
in Leighly et~al. 2007b) in which the \ion{O}{6} column density is $\sim$100 times the \ion{H}{1} column density 
(Tripp et al. 2008). Our inspection of the {\it HST} STIS spectrum of \phl\ also reveals \ion{N}{5} absorption at the 
same redshift, with $W_r$ an order of magnitude smaller than that of \ion{O}{6}; \ion{C}{4} absorption is not seen in a 
lower-resolution STIS spectrum. Associated \ion{O}{6} absorption is not rare, being seen in about 23\% 
(Frank et al. 2010) to 63\% (Tripp et al. 2008) of quasars. However, the $z=0.192186$ absorber in \phl\ is unusual, 
having the highest \ion{O}{6} column density and the highest \ion{O}{6}/\ion{H}{1} ratio among the 
14 associated \ion{O}{6} absorbers in Tripp et al. (2008). 
If this unusual \ion{O}{6} absorber is connected with the unusual 
\xray\ and UV properties of \phl, future spectroscopy should reveal
unusual associated \ion{O}{6} absorption in our \phl\ analogs as well.


In summary, the currently available data are unable to discriminate rigorously between the intrinsic \xray\
weakness and the \xray\ absorption scenarios. Occam's razor might initially suggest favoring the hypothesis that our 
\phl\ analogs are intrinsically \xray\ weak like \phl\ itself 
appears to be. However, the suggestion of heavy \xray\ absorption 
from the likely hard \xray\ spectra definitely needs further investigation. 
If our objects are indeed established to be heavily \xray\ absorbed, then
Occam's razor might alternatively be used to motivate reconsideration 
of the evidence that \phl\ itself is intrinsically \xray\ weak. 
The evidence for intrinsic \xray\ weakness presented by 
Leighly et~al. (2007a) appears strong (e.g., steep \xray\ spectrum 
with no apparent photoelectric absorption cut-off, \xray\ variability). 
Nevertheless, absorption effects in local AGNs are sometimes observed to 
be exceedingly complex, and perhaps some unusual absorbers 
or special absorption geometry
for this unusual quasar might yet be consistent with the observed 
\xray\ properties. 

\subsection{Correlations between \xray\ Weakness and UV Emission-Line Properties}\label{discuss:corr}
\begin{figure}[t]
    \centering
    \includegraphics[width=3.2in]{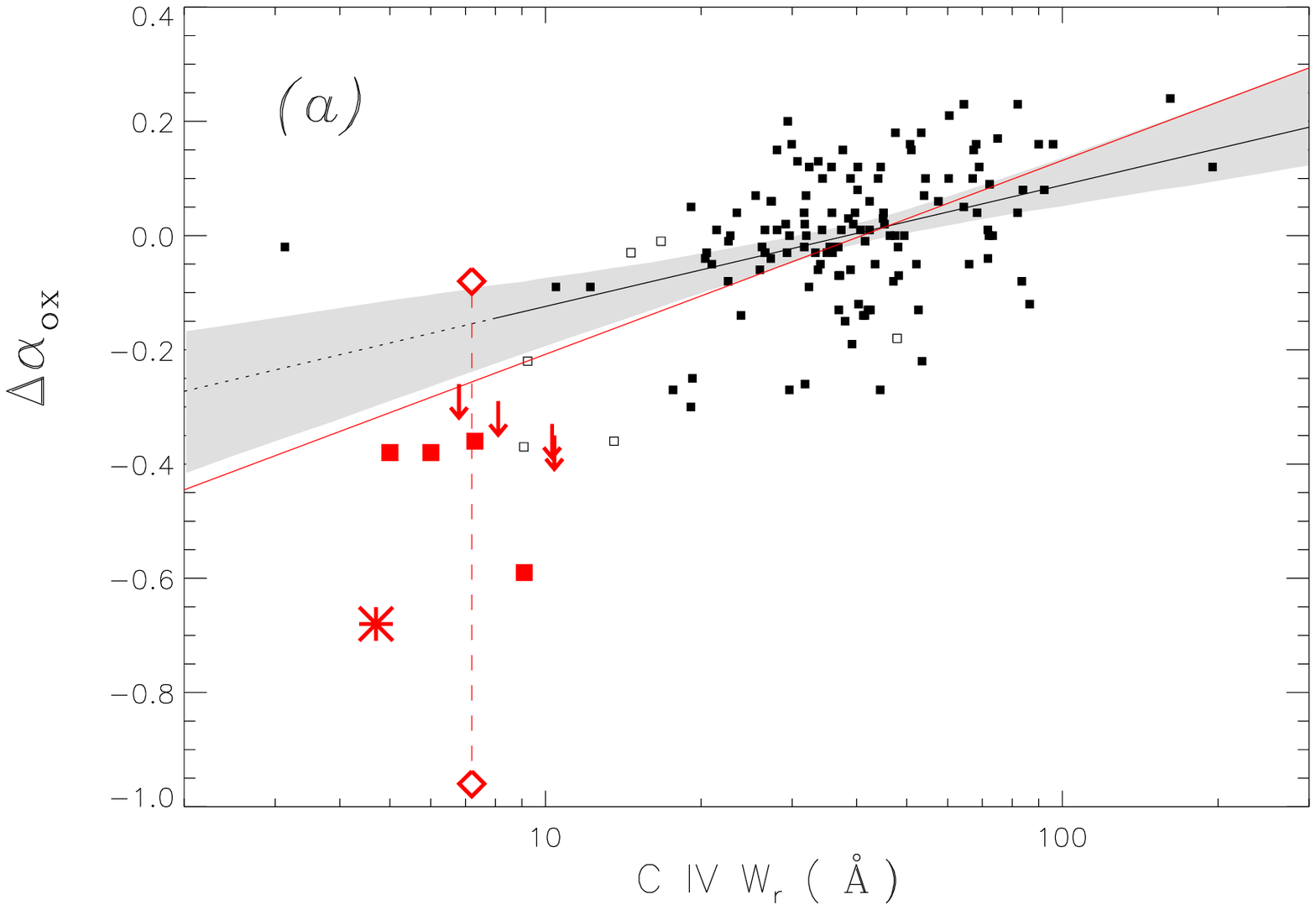}\\
    \includegraphics[width=3.2in]{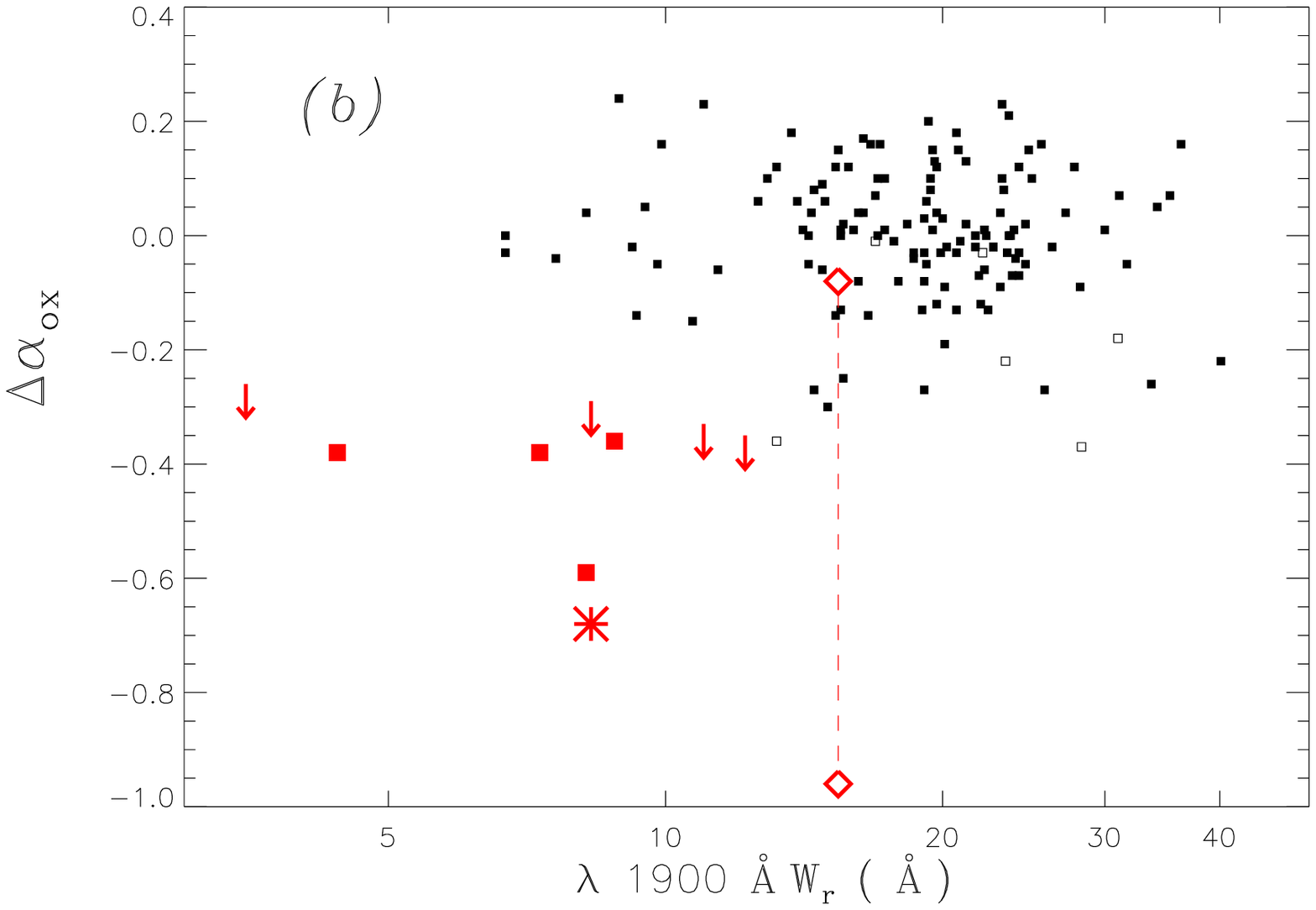}\\   
    \includegraphics[width=3.2in]{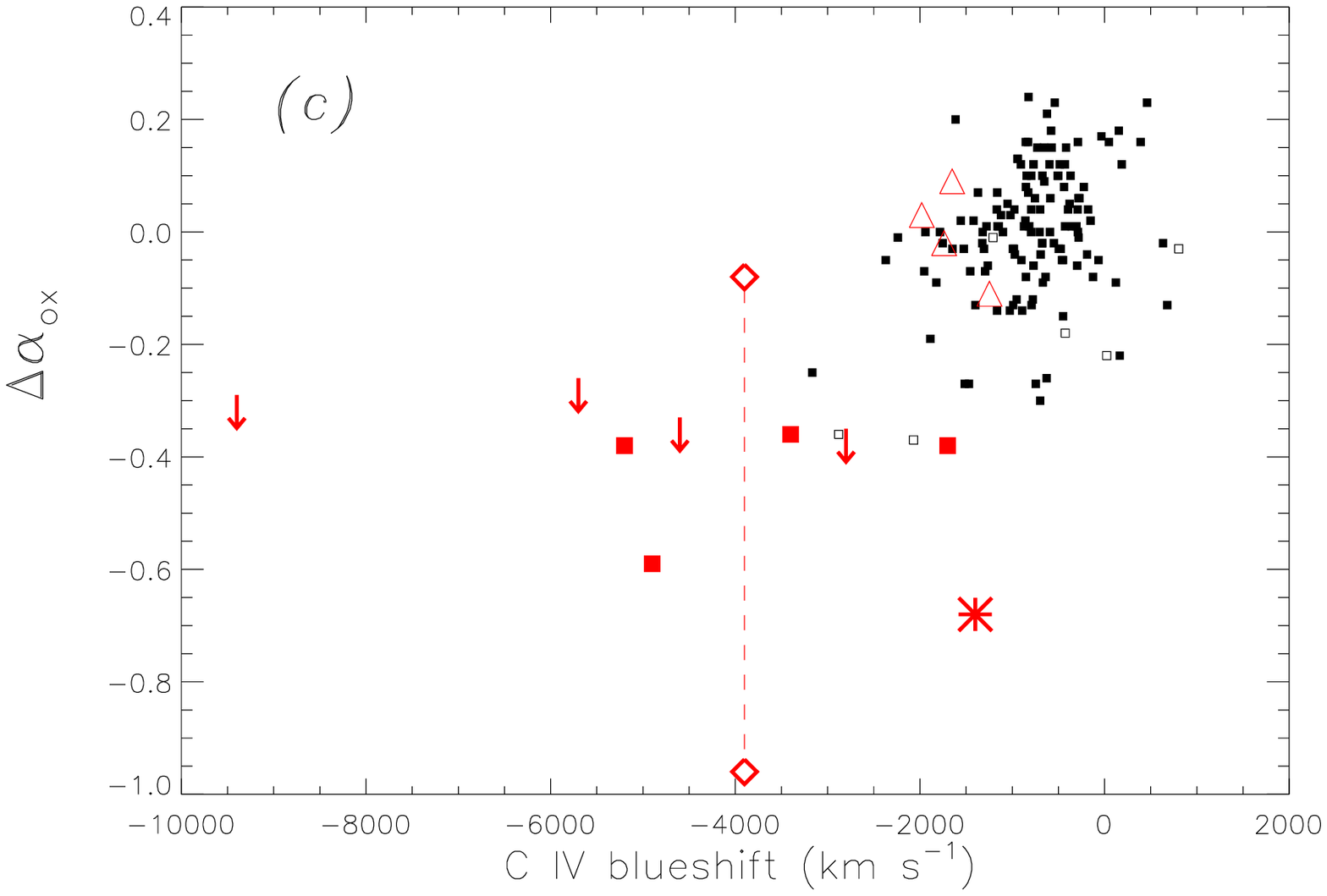}
    \caption{\footnotesize{The values of \daox\ plotted against the UV emission-line parameters of $W_r$(C~{\sc iv}) $(a)$,
             $W_r$($\lambda1900$~\AA) $(b)$, and C~{\sc iv} blueshift $(c)$. The black filled squares are \hbox{radio-quiet}, 
             \hbox{non-BAL} quasars in Sample~B of Gibson et~al. (2008a). The black open squares represent six sample B quasars that 
             have possible BAL or mini-BAL features based on visual inspection (see Footnote~16). All Sample~B quasars shown here 
             are \xray\ detected. 
             The red filled squares and downward arrows represent the \hbox{radio-quiet} \phl\ analogs detected and undetected in 
             \hbox{X-rays}, respectively. The red asterisk represents \phl. The red open diamonds represent the highly \xray\ 
             variable source PHL~1092 in two \xmm\ observations in 2003 (with greater \daox\ value) and 2008.
             The best-fit linear correlation for the 
             Sample~B quasars is shown by the black line in panel $(a)$; the solid part represents the range
             where Gibson et~al. (2008a) had statistically constraining data, while the dotted part is an extrapolation 
             of the solid line. The shaded area
             shows the 95\% confidence uncertainty range for this linear correlation determined with a nonparametric bootstrap
             method. The red solid line shows the best-fit correlation for a sample combining our \hbox{radio-quiet} \phl\ 
             analogs, \phl, and Sample~B quasars. The red open triangles in panel 
             $(c)$ show the four quasars from Gallagher et~al. (2005) having ``large'' C~{\sc iv} blueshift values.}
             \label{corr_fig}}
\end{figure}
We have quantitatively searched for relations between relative \xray\ weakness and the unusual UV \hbox{emission-line} 
properties. Correlation analyses were 
performed between \daox\ and the UV emission-line parameters listed in Table~\ref{qso_table}, 
using both Kendall's $\tau$ test and Spearman's \hbox{rank-order} analysis in the ASURV package. 
Kendall's $\tau$ test is preferred over Spearman's \hbox{rank-order} analysis in small-sample cases, 
where the sample size is $N\leq30$.\footnote{See Appendix~A3.3 of the ASURV manual at
http://astrostatistics.psu.edu/statcodes/asurv.} The results are presented in
Table~\ref{corr_table}. We found only a marginal correlation ($94.9\%$ correlation probability) 
between \daox\ and the blueshift of the 
\ion{C}{4} line when including both \hbox{radio-quiet} and \hbox{radio-intermediate} \phl\ analogs in our sample. 
The effective significance of this correlation is further weakened by the number of trials, since we have tested
for multiple correlations in Table~\ref{corr_table}. No other correlations were found; this is likely due to the small 
sample size and the limited range of parameter space spanned by the \phl\ analogs alone. If we use the
redshift values of Hewett \& Wild (2010) for our \phl\ analogs, the correlation between \daox\ and 
the \ion{C}{4} blueshift remains insignificant ($91.2\%$ correlation probability).

\begin{figure*}[t]
    \centering
    \includegraphics[width=5.5in]{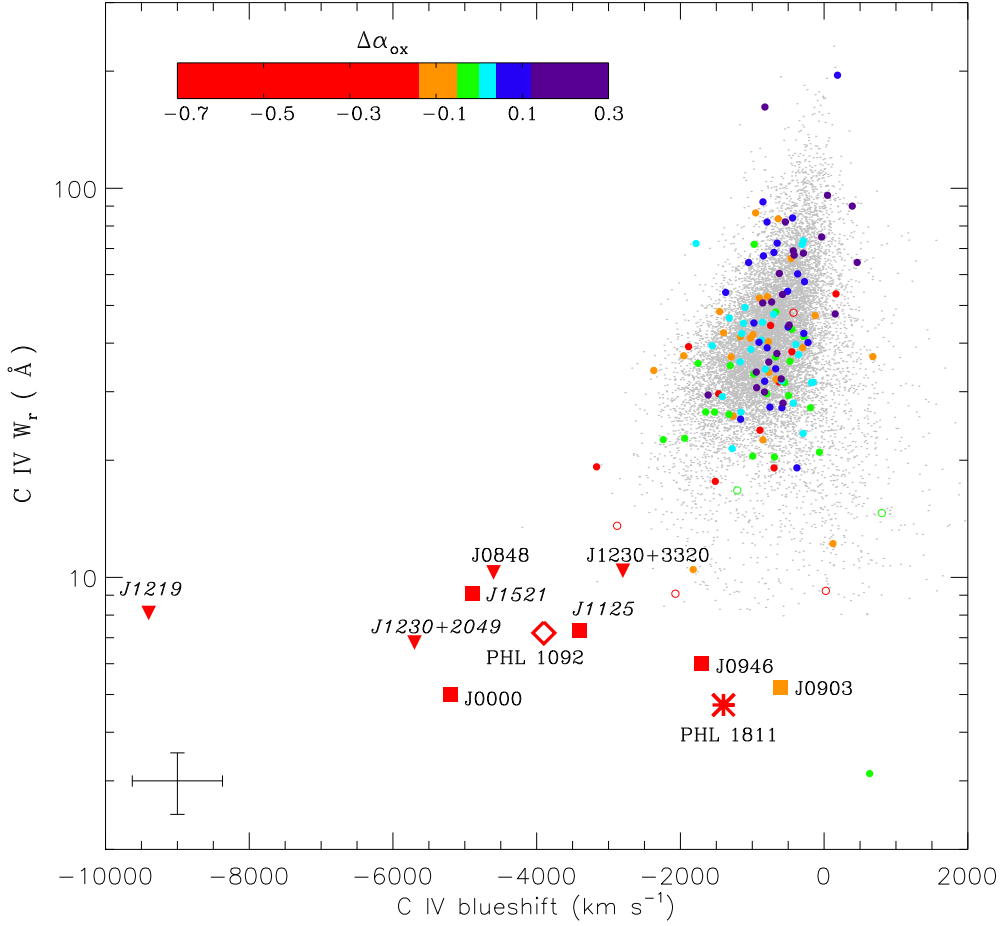}
    \caption{\footnotesize{The C~{\sc iv} blueshift plotted against $W_r$(C~{\sc iv}) for our \hbox{radio-quiet} \phl\ analogs (filled squares for
             \xray\ detected sources, filled upside-down triangles for \xray\ undetected sources), \phl\ (asterisk), and \hbox{radio-quiet}, \hbox{non-BAL}
             quasars in Sample~B of Gibson et~al. (2008a) (filled circles). The open circles represent six sample B quasars that have possible BAL or mini-BAL features based on visual inspection (see Footnote~16). These sources are color-coded according to their \daox\ values 
             (six color bins are used, each chosen to contain an approximately equal number of objects). The color bar shows the \daox\ 
             range for each color. Source names for \phl\ analogs are labelled in the format of 'J$hhmm$' for brevity, except for 
             J1230+3320 and J1230+2049. The sources with {\it italic} names are the \phl\ analogs for which we adopt different 
             redshift values other than those from the SDSS CAS. J0903+0708 is also shown using the line measurements from its
             HET spectrum. The red open diamond represents PHL~1092. The red color shows its \xray\ 
             weakness from the latest \xmm\ observation (see Miniutti et~al. 2009). The median errors for C~{\sc iv} blueshifts and 
             $W_r$(C~{\sc iv}) of 
             our \phl\ analogs are shown in the lower-left corner of the figure (the blueshift error includes contributions from both redshift 
             measurement and C~{\sc iv} wavelength measurement). The grey dots show the 13,582 \hbox{radio-quiet}
             quasars in Sample A of Richards et~al. (2010; see their Fig.~7).}
             \label{c4bew_fig}}
\end{figure*}

Fig.~\ref{corr_fig} shows \daox\ plotted against the UV emission-line parameters of $W_r$(\ion{C}{4}), 
$W_r$($\lambda1900$~\AA), and \ion{C}{4} blueshift for \hbox{radio-quiet} \phl\ analogs, \phl\ itself, and 
the \hbox{radio-quiet}, \hbox{non-BAL} quasars in Sample~B of Gibson et~al. (2008a). The addition of 
the Sample~B quasars and \phl\ increases the sample size to $N=148$ and broadens the parameter space 
of line measurements. The values of $W_r$(\ion{C}{4}) and \ion{C}{4} blueshift for the quasars in 
Sample~B are taken from Shen et~al. (2010); we modified the \ion{C}{4} blueshift values for the 
Sample~B quasars to account for the improved redshift measurements of
Hewett \& Wild (2010). The values of $W_r$($\lambda1900$~\AA) are taken from Gibson et~al. (2008a). 
PHL~1092 (see \S\ref{intro}) also has \phl-like UV emission-line properties and shows dramatic \xray\ 
variability.\footnote{We have inspected the 
available rest-frame UV spectra of PHL~1092 (Bergeron \& Kunth 1980; Leighly et~al. 2007b), covering Ly$\alpha$, 
Si~{\sc iv}, C~{\sc iv}, C~{\sc iii}], Fe~{\sc iii}~UV48, and Mg~{\sc ii}, and measured its UV emission-line parameters as 
listed in Table~\ref{qso_table}. We do not have spectral coverage of its UV Fe~{\sc ii} emission. 
PHL~1092 likely would have been selected as a \phl\ analog by our criteria in \S\ref{sample}, if it were in our redshift 
selection range of $2.125\leq z \leq2.385$ and were sufficiently bright.} We include this source in 
Fig.~\ref{corr_fig} and Fig.~\ref{c4bew_fig}, but not in the correlation analyses. 

A significant correlation between $W_r$(\ion{C}{4}) and \daox\ was reported by Gibson et~al. (2008a)
for their sample. 
After adding our \hbox{radio-quiet} \phl\ analogs and \phl, the correlation becomes more significant 
(see Table~\ref{corr_table} and the red solid line in Fig.~\ref{corr_fig}$a$). However, all of the \hbox{radio-quiet} \phl\ analogs 
and \phl\ itself lie well below the best-fit linear correlation (in log space) in Gibson et~al. (2008a) 
between $W_r$(\ion{C}{4}) and \daox. This remains true even if we consider the 95\% confidence uncertainty range shown in Fig.~\ref{corr_fig}$a$,
obtained using a nonparametric bootstrap method (Efron 1979). This deviation from the correlation found for typical 
\hbox{radio-quiet} quasars may indicate that the \hbox{radio-quiet} \phl\ analogs arise from a distinct population of
typical \hbox{radio-quiet} quasars. There is no significant correlation between 
$W_r$($\lambda1900$~\AA) and \daox\ (see Table~\ref{corr_table} and Fig.~\ref{corr_fig}$b$).

The \ion{C}{4} emission lines of quasars are generally known to have systematic blueshifts 
(e.g., Gaskell 1982; 
Richards et~al. 2010). We performed Spearman rank-order analysis and found a correlation between \ion{C}{4} blueshift and \daox\ 
for \hbox{radio-quiet} \phl\ analogs, \phl, and the Sample~B quasars. However, this correlation shows 
substantial scatter (see Fig.~\ref{corr_fig}$c$). Although the correlation probability \hbox{($>99.99\%$)} is formally high, 
the correlation coefficient ($r_S=0.39$) shows that this correlation is weaker than that between 
$W_r$(\ion{C}{4}) and \daox\ ($r_S=0.49$). Furthermore, it is difficult to determine if there is truly a single 
correlation or instead two populations artificially producing an apparent correlation. 

The red triangles in 
Fig.~\ref{corr_fig}$(c)$ show the four \hbox{radio-quiet}, \hbox{non-BAL} quasars from Gallagher et~al. (2005) which have
``large'' \ion{C}{4} blueshifts compared to most typical \hbox{radio-quiet}, \hbox{non-BAL} quasars. Their \xray\ brightnesses appear
to be normal. However, the magnitudes of their \ion{C}{4} blueshifts are \hbox{$\leq\ 2000$~km~s$^{-1}$}, while 
all but one of the \hbox{radio-quiet} \phl\ analogs in our sample have \ion{C}{4} blueshift magnitudes \hbox{$>2000$~km~s$^{-1}$}. 
The four sources from Gallagher et~al. (2005) are more similar to the typical \hbox{radio-quiet}, \hbox{non-BAL} quasars. Gallagher 
et~al. (2005) also suggested evidence for intrinsic absorption (at the level of $N_H\approx10^{22}$~cm$^{-2}$) for 
their ``large'' \ion{C}{4} blueshift objects via joint \xray\ spectral analyses. In order to search for any possible 
trend between intrinsic absorption and the \ion{C}{4} blueshift, we performed a band-ratio analysis for all the Sample~B 
quasars from Gibson et~al. (2008a). A harder \xray\ spectrum (with smaller effective photon index) could
indicate the presence of intrinsic absorption. However, no correlation was found between the effective photon 
indices and \ion{C}{4} blueshifts. 
\begin{figure*}[t]
    \centering
    \includegraphics[width=5.0in]{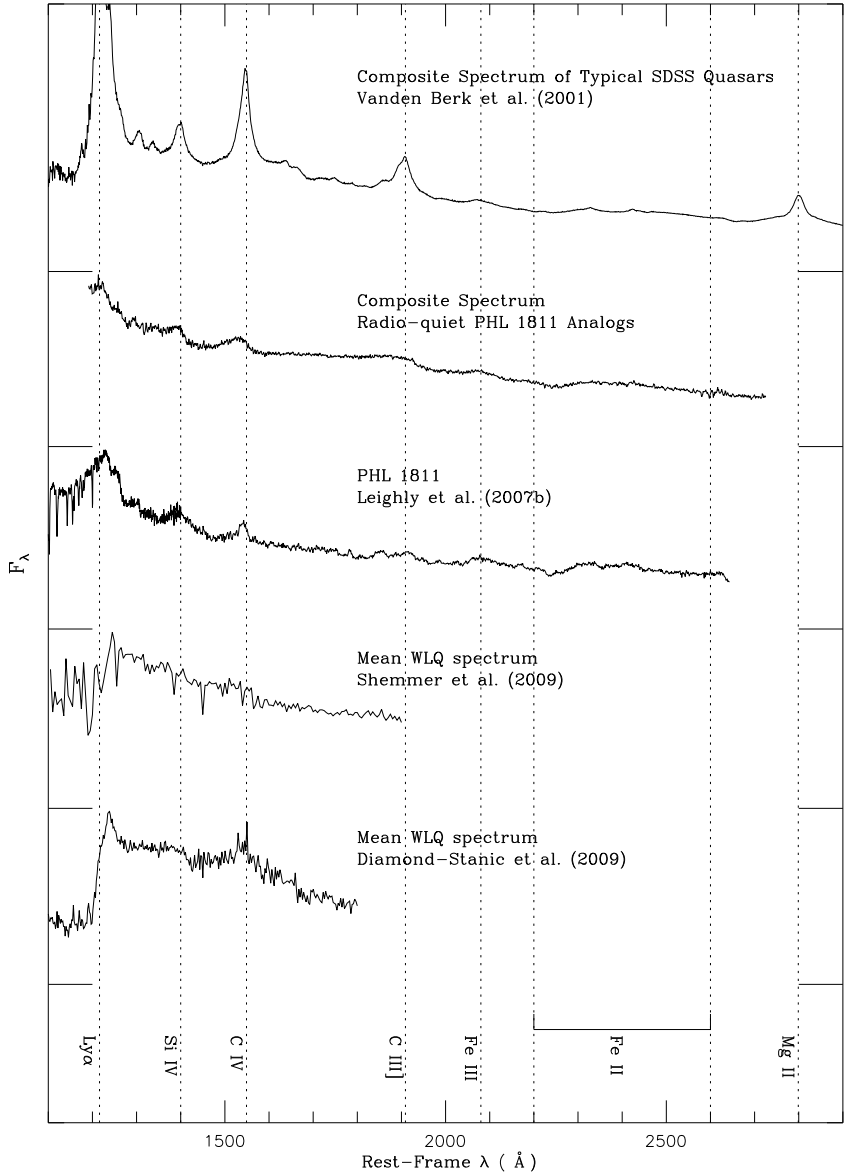}
    \caption{\footnotesize{Comparison of the median composite spectrum of \hbox{radio-quiet} \phl\ analogs with the mean 
             spectra of WLQs from Shemmer et~al. (2009), and Diamond-Stanic et~al. (2009). The $y$-coordinates are the 
             flux density ($F_\lambda$) in arbitrary linear units. The tick marks on the $y$-axis show the zero flux density 
             level for each normalized spectrum. The emission lines are labelled by the dotted vertical lines. Also shown are 
             the median composite spectrum of typical SDSS quasars from Vanden Berk et~al. (2001) and the spectrum of 
             \phl\ from the {\it Hubble Space Telescope} (Leighly et~al. 2007b).}
             \label{comp_fig}}
\end{figure*}

In Fig.~\ref{c4bew_fig}, the \ion{C}{4} blueshift is plotted against $W_r$(\ion{C}{4}) for our \hbox{radio-quiet} \phl\ analogs, 
\phl, and the Sample~B quasars in Gibson et~al. (2008a). The color for each source shows its \daox\ value, indicating its 
relative \xray\ brightness. The grey dots represent the 13,582 \hbox{radio-quiet} Sample A quasars from Richards et~al. (2010) 
which do not generally have constraining information on their \xray\ properties. Motivated by much past work, Richards 
et~al. (2010) described a BELR model with both ``disk'' and ``wind'' components. The \ion{C}{4} blueshift and $W_r$(\ion{C}{4})
show the trade-off between these two components, which ultimately depends upon the shape of the ionizing continuum. The
claimed interpretation is that quasar BELRs change from disk-dominated (with a more ionizing SED) to wind-dominated 
(with a less ionizing SED) as a quasar's location moves from the upper right corner to the lower left corner of 
Fig.~\ref{c4bew_fig}. The relative \xray\ brightness becomes moderately weaker for the Sample~B quasars in Gibson et~al. 
(2008a) following this trend. Our \hbox{radio-quiet} \phl\ analogs show extreme behavior in the \daox$-$\ion{C}{4} 
blueshift$-$ $W_r$(\ion{C}{4}) parameter space; note they are essentially disjoint from the main quasar population. 
This suggests they may have extreme wind-dominated BELRs. 

\subsection{The Relation Between \phl\ Analogs and Weak-Line Quasars}\label{discuss:wlq}

The SDSS has discovered $\sim80$ high-redshift ($z>2.2$) quasars with extremely weak or undetectable UV emission lines 
(WLQs; e.g., Diamond-Stanic et~al. 2009); similar objects also likely exist at lower redshifts
(e.g., Plotkin et~al. 2010b). 
In this section, we discuss the observational relation between our \phl\ analogs and WLQs more generally.  WLQs are 
defined as quasars with $W_r$(Ly$\alpha$+\ion{N}{5})$<$15.4\,\AA, measured between 1160\,\AA\ and 1290\,\AA\ in the rest 
frame (Diamond-Stanic et~al. 2009).  Measurements of Ly$\alpha$ $W_r$ are difficult with our wavelength coverage, but many 
of our \phl\ analogs likely meet the WLQ definition; for example, J0946+2744 was identified as a WLQ by Shemmer et~al. 
(2009).
\begin{figure*}[t]
    \centering
    \includegraphics[width=5.0in]{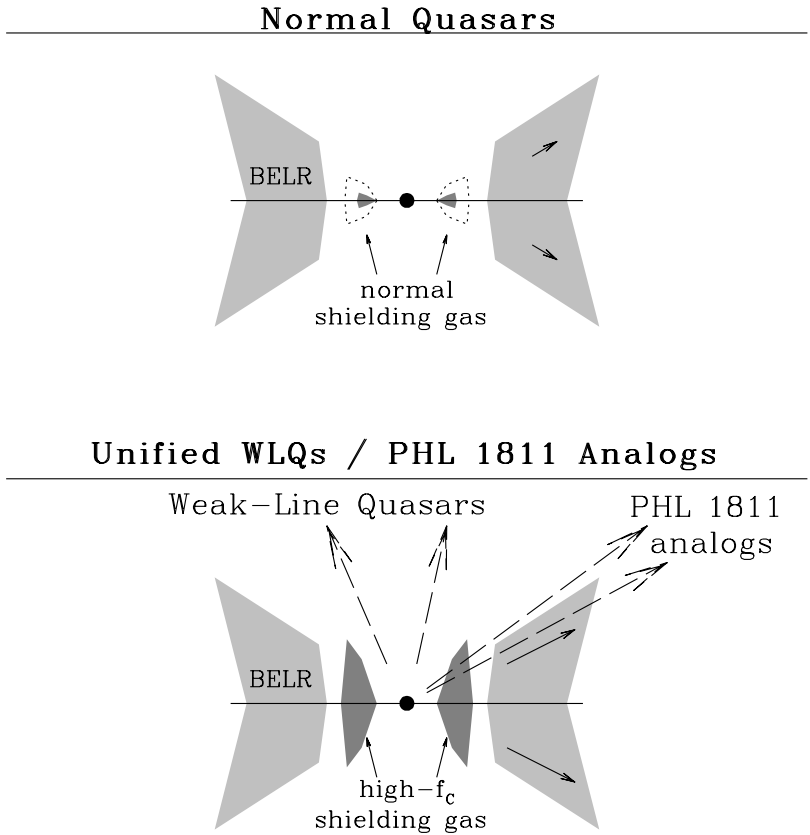}
    \caption{\footnotesize{Cartoon illustration of the WLQ/\phl\ analog unification hypothesis.
             Each panel is a side view of an accretion disk around a black hole, showing the
             \xray\ and UV continuum sources (central black dot), shielding gas (grey), and broad
             emission line region (BELR; light grey).  The BELR is shown as shaded regions for convenience, but likely consists of ``clouds''.
             Normally (top panel), the shielding gas only covers part of the BELR, resulting
             in moderate emission-line blueshifts at best (short arrows in the BELR). The dotted outline illustrates the fact that normal 
             quasars are expected to have a range of shielding gas covering factors ($f_{\rm C}$) and columns (and thus a range of wind strengths). 
             When the shielding gas has a BELR covering factor of $f_{\rm C}\gtrsim$~80\% (bottom panel), a strong wind
             is generated (long arrows in the BELR).  When such a quasar is viewed through
             the shielding gas and thus close to the wind direction, a \phl\ analog is seen
             (\xray\ absorbed, high blueshift).  When such a quasar is viewed away from the
             wind direction, a weak-line quasar is seen (\xray\ normal, low blueshift).}
             \label{unify_fig}}
\end{figure*}

There has been some past speculation about a possible connection between \phl\ itself and WLQs (e.g., Leighly et~al. 2007b; 
Shemmer et~al. 2009), but we here have the advantage of having a carefully selected sample of \phl\ analogs with established 
basic \xray\ properties. 
Fig.~\ref{comp_fig} compares the median composite spectrum of radio-quiet \phl\ analogs with the mean spectra of WLQs from 
Shemmer et~al. (2009) and Diamond-Stanic et~al. (2009).\footnote{Because our \phl\ analogs
all lie in a similar part of the $W_r$(C~{\sc iv}) vs. C~{\sc iv} blueshift parameter space, their composite spectrum should be
physically meaningful at least to first order (cf. \S5 of Richards et~al. 2010).}  
\phl\ analogs and WLQs both have weak Ly$\alpha$, \ion{Si}{4}, and 
\ion{C}{4} emission lines. \phl\ analogs have prominent UV \ion{Fe}{2} and \ion{Fe}{3} emission, while most WLQ samples have 
been selected at high redshift and generally lacked spectral coverage of UV Fe emission.  However, inspection of the 
lower-redshift WLQ samples of Plotkin et~al. (2010ab) shows that their UV Fe emission can be just as prominent 
relative to the broad lines as it is in \phl\ analogs. Furthermore, the two WLQs studied by Shemmer et al. (2010) both have 
prominent optical Fe II, similar to PHL 1811 itself (Leighly et al. 2007b).

When defining our sample of \phl\ analogs (see \S\ref{sample:select}), we first selected 54 objects with weak high-ionization 
lines, 32 of which can be considered to be \hbox{non-BAL} WLQs. 29 of the 32 \hbox{non-BAL} WLQs are radio quiet. Eight of 
these 29 sources having strong UV Fe emission and \ion{C}{4} blueshifts were retained as our radio-quiet \phl\ analogs. At 
least from an observational point of view, \phl\ analogs thus appear to be a subset ($\simeq 30\%$) of WLQs. 
Our radio-quiet \phl\ analogs are \xray\ weak 
compared to typical quasars, while WLQs are generally not \xray\ weak as a population (Shemmer et~al. 2009). However, there 
are a few known \xray\ weak WLQs (e.g., J1302+0030, J1421+3433, and perhaps J1335+3533 and J1532$-$0039 in Shemmer et~al. 
2006, 2009) which may be \phl\ analogs. We do not have reliable measurements of the \ion{C}{4} blueshifts for these sources 
because of either the wavelength coverage or the quality of their SDSS spectra. Near-infrared spectroscopy for many WLQs 
(e.g., Shemmer et~al. 2010) is required to assess if the \xray\ weak WLQs have notably strong UV \ion{Fe}{2} and \ion{Fe}{3} 
emission.

While the observational relation between \phl\ analogs and WLQs appears fairly simple, as described above, any physical 
relation between these two classes may be much more complex.  The \xray\ weakness and distinctive emission lines of 
\phl\ analogs may ultimately be due to some extreme physical parameter (or parameters), such as accretion rate 
(see \S\ref{intro}). In contrast, at least some WLQs appear simply to have anemic broad line regions (Shemmer et~al. 2010), 
and some may be in a specific evolutionary stage where the quasar activity has only recently begun (Hryniewicz et~al. 2010).

However, there is another scenario which might unify \phl\ analogs and most WLQs, which we illustrate in Fig.~\ref{unify_fig}. 
As mentioned in \S\ref{discuss:intrinsic}, suppose there is a subset of quasars in which the shielding gas covers all or most 
of the BELR (which likely consists of ``clouds''), but little more than the BELR. 
A sufficient column of such gas could absorb \xray\ and other ionizing photons before they reach the BELR, resulting in weak 
\hbox{broad-line} emission and wind acceleration without overionization. (In normal quasars with strong \hbox{broad-line} 
emission, a range of shielding gas covering factors and columns are likely to exist, but the majority of the BELR gas cannot 
be shielded by such high columns and must be exposed to the ionizing continuum.) 
When such quasars are observed {\it through the shielding gas}, a \phl\ analog would be seen; 
when they are observed from other directions (where shielding gas is unlikely to be seen), 
an \xray\ normal WLQ would be seen.
A \phl\ analog fraction of 30\% among WLQs is consistent with the estimated BELR covering fraction (e.g., Maiolino et~al. 2001).  
The low \ion{C}{4} $W_r$ values in \phl\ analogs require that $\gtrsim$~80\%$\pm$20\% of the BELR be covered by shielding 
gas.\footnote{We 
have illustrated this using shielding gas of variable $f_{\rm C}$ in Fig.~\ref{unify_fig}.  Alternatively, as depicted
in Fig.~15 of Leighly (2004), an \xray\ source of varying height above the disk could result in varying fractions of the BELR 
being illuminated by \hbox{X-rays} (see also Miniutti \& Fabian 2004).}
As coverage of randomly distributed BELR clouds by shielding gas with the same covering factor would be physically implausible, 
this scenario requires the BELR in \phl\ analogs and WLQs to have a \hbox{non-random} geometrical distribution (see, e.g., 
\S4.3 of Risaliti et~al. 2011 and references therein). The large \ion{C}{4} blueshifts of \phl\ analogs could be produced 
in lines of sight into an accelerating wind, for example, from a rotating, \hbox{disk-like} BELR (Murray et~al. 1995). 

This simple
unification scenario makes several predictions. The \ion{O}{6} absorption in \phl\ (see \S\ref{discuss:intrinsic}) 
is evidence for the existence of highly ionized \hbox{UV-absorbing} gas along our sightline to its central 
engine; \phl\ analogs might have an excess of such absorption, but WLQs should not.
Broad emission line $W_r$ values in WLQs and \phl\ analogs should be comparable, consistent with 
the limited data available to date (e.g., the Appendix of this work and Shemmer et~al. 2010).
\hbox{Narrow-line} emission {\it could} be seen in WLQs and \phl\ analogs in this model, but is unlikely given the 
observed \hbox{anti-correlation} (Boroson \& Green 1992; Netzer et~al. 2004) of [\ion{O}{3}] strength with the 
\ion{Fe}{2}/H$\beta$ ratio,\footnote{Fe~{\sc ii}/H$\beta$ is defined as the ratio between $W_r$(Fe~{\sc ii}) in 
4434--4684~\AA\ and $W_r$(H$\beta$).} which is large in \phl\ (Leighly et~al. 2007b), J1521+5202 (the Appendix), and 
at least two WLQs (Shemmer et~al. 2010). Additional observations to test these predictions are needed.

\subsection{Radio-Intermediate \phl\ Analogs}\label{discuss:radio}
\begin{figure}[t]
    \centering
    \includegraphics[width=3.5in]{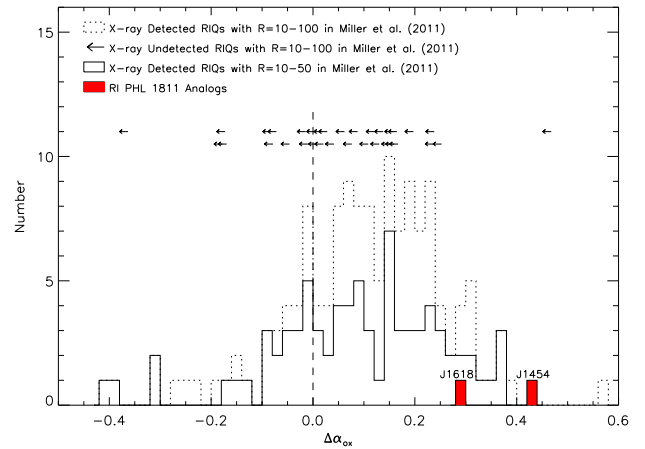}
    \caption{\footnotesize{Comparison of the \daox\ value distributions for our \hbox{radio-intermediate} \phl\ analogs and typical 
             \hbox{radio-intermediate} quasars in the primary sample of Miller et~al. (2011). The red histogram represents our 
             \hbox{radio-intermediate} \phl\ analogs, with their names labeled in the format of 'J$hhmm$' for brevity. The solid 
             histogram shows the \daox\ value distribution of \xray\ detected typical \hbox{radio-intermediate} quasars with 
             similar \hbox{radio-loudness} values ($R=10$--$50$) to our \phl\ analogs, while the dotted histogram is for the 
             whole sample of \xray\ detected \hbox{radio-intermediate} quasars in Miller et~al. (2011). The leftward arrows are 
             for the \xray\ undetected \hbox{radio-intermediate} quasars in Miller et~al. (2011). The dashed vertical line shows \daox $=0$.}
             \label{ricom_fig}}
\end{figure}
The two \hbox{radio-intermediate} sources in our sample, J1454+0324 and J1618+0704, are \xray\ bright with \daox$=0.42$ and
\daox$=0.28$, respectively. \hbox{Radio-intermediate} sources probably have significant relativistic jets, which often have associated \xray\
emission (e.g., Worrall et~al. 1987; Miller et~al. 2011). Fig.~\ref{ricom_fig} compares the \xray\ ``excesses'' in these two \phl\ 
analogs (assessed with \daox) with those of typical \hbox{radio-intermediate} quasars from the primary sample of 
Miller et~al. (2011). The \daox\ values of our two \hbox{radio-intermediate} \phl\ analogs are relatively large compared to those 
of typical \hbox{radio-intermediate} quasars with similar radio loudness values ($R=10$--$50$), and also compared to
the whole \hbox{radio-intermediate} quasar sample ($R=10$--$100$; also see Fig. 7 of Miller et~al. 2011). Note that the 
radio-loudness values in Miller et~al. (2011) have been converted to our definition (they used 
$f_{2500\mbox{\rm~\scriptsize\AA}}$ for the optical flux density). 

We performed \xray\ spectral analyses for the two \hbox{radio-intermediate} \phl\ analogs, since both of them have sufficient counts
for basic spectral fitting. The \xray\ spectra were extracted with the {\sc psextract} routine in CIAO v4.2, 
using a $3''$ radius aperture centered on the \xray\ position for each source. Background spectra were 
extracted using annular regions with inner radii of $6''$ and outer radii of $9''$. Both background regions are 
free of \xray\ sources. Spectral fitting was executed using XSPEC v12.5.1 (Arnaud 1996). The \xray\ spectra were grouped to 
have at least 20 (10) counts per bin for J1454+0324 (J1618+0704). We used 
a \pl\ model with a Galactic absorption component, in which the Galactic column density is fixed to the values from 
Table~\ref{aox_table}. We also tried another model similar to the first, but 
adding an intrinsic neutral absorption component. Table~\ref{fit_table} shows the \xray\ spectral fitting results. 
The quoted errors or the upper limits for the best-fit parameters are at 90\% confidence for one parameter of 
interest ($\Delta \chi^2=2.71$; Avni 1976). For both sources, we found no evidence for strong intrinsic 
neutral absorption; adding an intrinsic neutral absorption component did not improve the fit quality. The best-fit
photon indices for both sources are consistent with those from band-ratio analysis (see the last column of Table~\ref{cts_table}).

J1618+0704, which has moderate radio emission ($R=35$), has a relatively flat \xray\ \pl\ continuum ($\Gamma=1.48^{+0.29}_{-0.28}$) 
indicating that its \xray\ flux probably has a substantial contribution from relativistic jets (e.g., Wilkes \& Elvis 1987;
Page et~al. 2005). The \xray\ brightness and \xray\ spectral shape of J1618+0704 suggest 
that the \xray\ emission associated with jets has not been significantly diminished, even if the \xray\ emission from the
ADC has been absorbed or quenched 
as for the \hbox{radio-quiet} \phl\ analogs (i.e., the observed spectrum may be dominated by jet-linked emission). 

The other \hbox{radio-intermediate} \phl\ analog, J1454+0324, has a steeper \xray\ continuum ($\Gamma=2.13^{+0.18}_{-0.17}$).
The radio-loudness value \hbox{($R=12.8$)} of this source is just above the upper limit for \hbox{radio-quiet} sources. However, 
J1454+0324 is brighter in \hbox{X-rays} than most typical \hbox{radio-intermediate} quasars with significantly higher $R$ values 
(see Fig.~\ref{ricom_fig}). The notable \xray\ brightness and the steep \xray\ continuum indicate that this source may have 
a different physical nature from that of J1618+0704. One possibility is that J1454+0324 may be similar to 
BL~Lac objects. Although BL~Lac objects are generally radio-loud, it is possible that the BL~Lac population has a small
radio-faint tail (e.g., Plotkin et~al. 2010b). In this scenario, J1454+0324 would have a relatively featureless UV spectrum 
because its emission lines have been somewhat diluted by a relativistically boosted UV continuum (though not enough to 
make it a {\it bona~fide} BL~Lac object), which allows it to pass our criteria for selecting \phl\ analogs. 
The \aox\ value for J1454+0324 is consistent with those for the majority of 
BL~Lac objects (e.g. Shemmer et~al. 2009; Plotkin et~al. 2010a) for which the \xray\ emission is relativistically boosted. 
The \xray\ spectral slope of J1454+0324 is also consistent with those of BL~Lac objects (e.g., Donato et~al. 2005). 
However, there are several problems with a BL~Lac-like interpretation. First, J1454+0324 has relatively strong 
observed \ion{Fe}{2} and \ion{Fe}{3} emission, which should also be diluted by a relativistically boosted continuum. Second,   
J1454+0324 has a blue UV/optical continuum ($\alpha_\nu=-0.28$), while typical BL~Lac objects have red continua
(e.g., Stein et~al. 1976). Third, there is no significant variability (only $\approx3\%$ level) between the two SDSS 
spectroscopic epochs of J1454+0324. Future observations (e.g., polarization measurements) of J1454+0324 in the UV/optical band 
may further constrain its nature which presently remains uncertain. 

\section{Summary and Future Studies}\label{summary}

We report on the \xray\ properties of 10 quasars identified by the SDSS at high redshift \hbox{($z\approx2.2$)}. 
Seven of them have new \chandra\ observations, while three have archival \chandra\ or \xmm\ data. This sample of 10 quasars
(eight \hbox{radio-quiet}, two \hbox{radio-intermediate}) was selected to have unusual UV emission-line properties similar to those 
of \phl, a confirmed intrinsically \xray\ weak quasar.
Our main results are the following:

\begin{enumerate}
\item Four of the eight \hbox{radio-quiet} \phl\ analogs are detected in \hbox{X-rays}. The distribution of \daox\ values
for \hbox{radio-quiet} \phl\ analogs is substantially different from that of typical \hbox{radio-quiet}, \hbox{non-BAL} quasars. 
All of the eight \hbox{radio-quiet} \phl\ analogs, without exception, are anomalously \xray\ weak by a mean factor of $\approx13$. 
\item The fraction of \hbox{radio-quiet} \phl\ analogs in the total \hbox{radio-quiet}, \hbox{non-BAL} quasar population is estimated
to be $\lesssim1.2\%$, and within the range of $0.3$--$2.8\%$ (at the 90\% confidence level). The small fraction
of \phl\ analogs does not present material difficulties to the utility of \xray\ surveys for finding AGNs. 
\item The currently available data are unable to discriminate rigorously between the intrinsic \xray\ weakness and
heavy \xray\ absorption scenarios. The \hbox{radio-quiet} \phl\ analogs have blue UV/optical continua 
without detectable broad absorption lines or significant dust reddening, supporting the hypothesis
that they are intrinsically \xray\ weak like \phl\ itself. Their unusual UV emission-line 
properties can also be naturally explained by a model with an intrinsically \xray\ weak SED. However, a stacking 
analysis of the \hbox{radio-quiet} \phl\ analogs shows evidence for a hard \xray\ spectrum (with large error bars), 
which could be due to the presence of \xray\ absorption. In this scenario, the \xray\ absorption must occur on a scale
smaller than the BELR and must cover most of the BELR. It is possible
that very high-ionization (e.g., \ion{O}{6}) absorbers could be connected to the unusual \xray\ and UV properties of 
\phl\ analogs and \phl\ itself. 
\item Our sample of \hbox{radio-quiet} \phl\ analogs empirically supports the connection between soft SEDs (\xray\ weak and 
UV/optical strong) and \phl-like UV emission lines.
\item We have investigated correlations between relative \xray\ brightness and UV emission-line properties 
[e.g., $W_r$(\ion{C}{4}) and \ion{C}{4} blueshift] for a sample combining our \hbox{radio-quiet} \phl\
analogs, \phl, and typical type 1 quasars. A significant correlation is found between $W_r$(\ion{C}{4}) and 
\daox. However, our \hbox{radio-quiet} \phl\ analogs show a notable deviation from the best-fit power-law correlation for typical 
type 1 quasars. The behavior of \phl\ analogs in the \daox$-$\ion{C}{4} blueshift$-$$W_r$(\ion{C}{4}) 
parameter space suggests that they may have extreme wind-dominated broad 
emission-line regions and/or may be a distinct population from typical \hbox{radio-quiet} quasars.
\item From an observational point of view, \phl\ analogs appear to be a subset ($\simeq 30$\%) of WLQs that are \xray\ weak and have 
strong UV Fe emission and \ion{C}{4} blueshifts. 
The existence of a subset of quasars in which a large column of \hbox{high-ionization} shielding gas
exists along all or most of the sightlines to the BELR, but along very few
other sightlines, could potentially unify the PHL 1811 analogs and WLQs.
\item The two \hbox{radio-intermediate} \phl\ analogs are \xray\ bright. One (J1618+0704) has an \xray\ spectrum 
consistent with jet-dominated \xray\ emission, and the other (J1454+0324) is more similar to BL~Lac objects in terms of
its \xray\ brightness and spectral shape. However, for J1454+0324, the strong UV Fe emission, blue UV/optical continuum, 
and lack of variability do not support a BL~Lac-like interpretation. 
\end{enumerate}

The observed connection between \phl-like UV emission lines and soft SEDs appears to provide a practical 
and economical way to find \xray\ weak quasars. 
\xray\ observations of further targets selected using \hbox{high-quality} UV/optical spectroscopy from large-area 
surveys, such as SDSS-III (Eisenstein et~al. 2011), LAMOST
(The Large Sky Area Multi-Object Fiber Spectroscopic Telescope; Su et~al. 1998; Wu et~al. 2010b), 
and BigBOSS (Schlegel et~al. 2009), should identify many more \phl\ analogs that are \xray\ weak. 

Further \xray\ studies of \phl\ analogs would be helpful to clarify their nature. 
Four of our eight \hbox{radio-quiet} \phl\ analogs are undetected in \hbox{X-rays}, leaving only upper limits
on their \xray\ fluxes. 
Deeper \xray\ observations are required to convert the individual upper limits into detections and clarify
the distribution of \xray\ weakness for \phl\ analogs. A clear bimodality in the distribution of the relative 
\xray\ brightness (\daox) for a sample combining typical type~1 quasars and \phl\ analogs would suggest that \phl\ analogs
are indeed a distinct population of quasars; alternatively, the lack of such bimodality would suggest that \phl\ analogs 
simply represent one extreme of a continuum of quasar properties. 
Furthermore, our current sample of \phl\ analogs suffers from limited sample size largely because of the restricted redshift range
used in target selection (see Fig.~\ref{zMi_fig}).  Selection over a broader range of redshift will allow us to investigate 
additional optically bright objects. An improved sample size will better characterize general \xray\ properties and allow more reliable 
correlation tests with emission-line properties.
\xray\ spectroscopy should be able to test directly if \phl\ analogs are intrinsically \xray\ weak; extensive \xray\
observations provided strong evidence that the \xray\ weakness of \phl\ was intrinsic rather than due to absorption (Leighly et~al. 
2007a). A stacking analysis suggests the average \xray\ spectrum of \hbox{radio-quiet} \phl\ analogs is harder than those
of typical quasars, which may indicate the presence of heavy \xray\ absorption. 
Stacking analyses with deeper and/or more observations can test and extend this result with better photon statistics. 
If the \hbox{X-rays} are absorbed, we expect to see a photoelectric absorption cutoff in high-quality \xray\ spectra, which can
rigorously distinguish the intrinsic \xray\ weakness and the heavy \xray\ absorption scenarios.  
\xray\ spectral measurements of the photon index of the hard \xray\ power law can also constrain $L/L_{\rm Edd}$ 
values (e.g., Shemmer et~al. 2008), providing insight as to whether high  $L/L_{\rm Edd}$ is plausibly the ultimate
cause of their remarkable SEDs and emission-line properties (see \S\ref{intro}). 
Given the low \xray\ fluxes \hbox{($\approx10^{-15}$~erg~cm$^{-2}$~s$^{-1}$} from $0.5$--$2.0$~keV) of the \hbox{radio-quiet} 
\phl\ analogs, 
they will be excellent targets for future missions, e.g., the {\it International X-ray Observatory}
({\it IXO}; e.g., White et~al. 2010), which will have far superior \xray\ spectroscopic capability. 
Long-term \xray\ monitoring of our \phl\ analogs will test whether they 
have state transitions like PHL~1092, and will help reveal whether instabilities of 
\xray\ emitting ADCs universally exist in \phl\ analogs. 

Finally, comparisons of high-quality spectra of PHL 1811 analogs and WLQs 
will help pin down any connection between them.  UV spectroscopy of
our PHL 1811 analogs covering the \ion{O}{6} line will reveal what fraction of
them have very highly ionized UV absorption near the quasar redshift.  
Near-infrared spectroscopy of our \phl\ analogs covering the H$\beta$ region should allow estimation of their
$M_{\rm BH}$ and $L/L_{\rm Edd}$ values via the standard virial method
(e.g., Vestergaard \& Peterson 2006; Shen et~al. 2008), again testing for  
high $L/L_{\rm Edd}$.\footnote{We note that application of the standard
virial method for SMBH mass and $L/L_{\rm Edd}$ determination will require caution, since extreme objects such as 
\phl\ analogs may not obey standard virial relations (cf. \S6.4 of Richards et~al. 2010). Comparison of $L/L_{\rm Edd}$
values from the \xray\ continuum based method and the virial method will thus be valuable as a consistency check.} 
Such spectroscopy will also constrain their [\ion{O}{3}] narrow-line regions, allowing comparisons with 
eigenvector~1 of Boroson \& Green (1992). 
See the Appendix for near-infrared spectroscopy of J1521+5202. Near-infrared spectroscopy of additional 
\hbox{high-redshift} WLQs, covering the \ion{Fe}{2} and \ion{Fe}{3} transitions, will also reveal if the 
\xray\ weak subset of this population is made up of \phl\ analogs (see \S\ref{discuss:wlq}).


\begin{acknowledgments}

We thank the anonymous referee for providing helpful comments. 
We thank A.~C.~Fabian, E.~D.~Feigelson, and B.~P.~Miller for helpful discussions. We acknowledge financial 
support from Chandra \xray\ Center grant GO0-11010X (J.W., W.N.B.), NASA ADP 
grant NNX10AC99G (J.W., W.N.B.), and NSERC (P.B.H.).

Funding for the SDSS and SDSS-II has been provided by the Alfred P. Sloan Foundation, 
the Participating Institutions, the National Science Foundation, the U.S. Department 
of Energy, the National Aeronautics and Space Administration, the Japanese Monbukagakusho, 
the Max Planck Society, and the Higher Education Funding Council for England. The SDSS 
Web site is http://www.sdss.org/. The Hobby-Eberly Telescope (HET) is a joint project of 
the University of Texas at Austin, the Pennsylvania State University, Stanford University, 
Ludwig-Maximilians-Universität München, and Georg-August-Universität Göttingen. The HET is 
named in honor of its principal benefactors, William~P.~Hobby and Robert~E.~Eberly. 
The Apache Point Observatory 3.5-meter telescope is owned and operated by the Astrophysical Research Consortium.

\end{acknowledgments}


\appendix
\section{Near-Infrared Spectroscopy of J1521+5202}
\begin{figure*}[t]
    \figurenum{A1}
    \centering
    \includegraphics[width=3.5in]{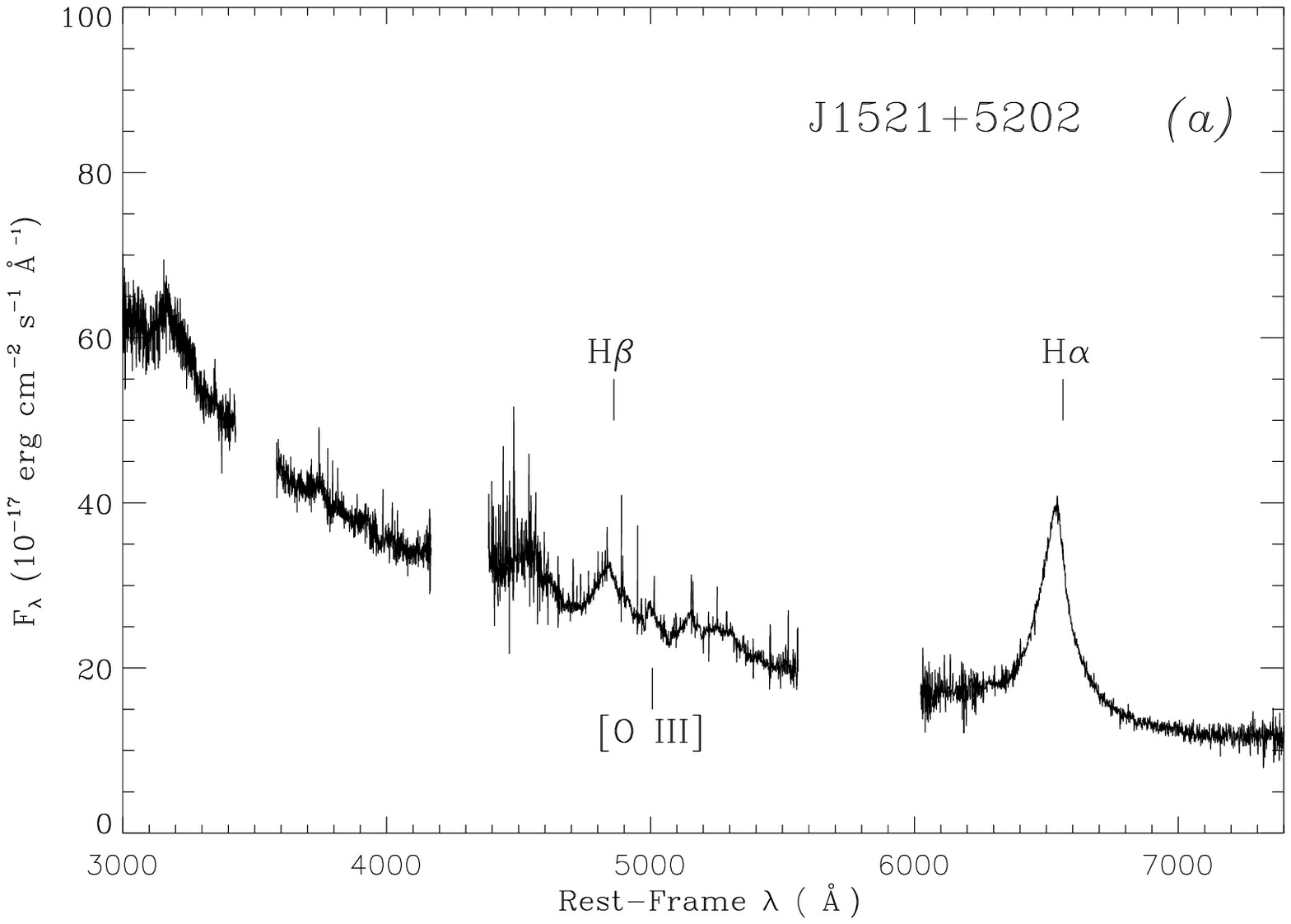}
    \includegraphics[width=3.5in]{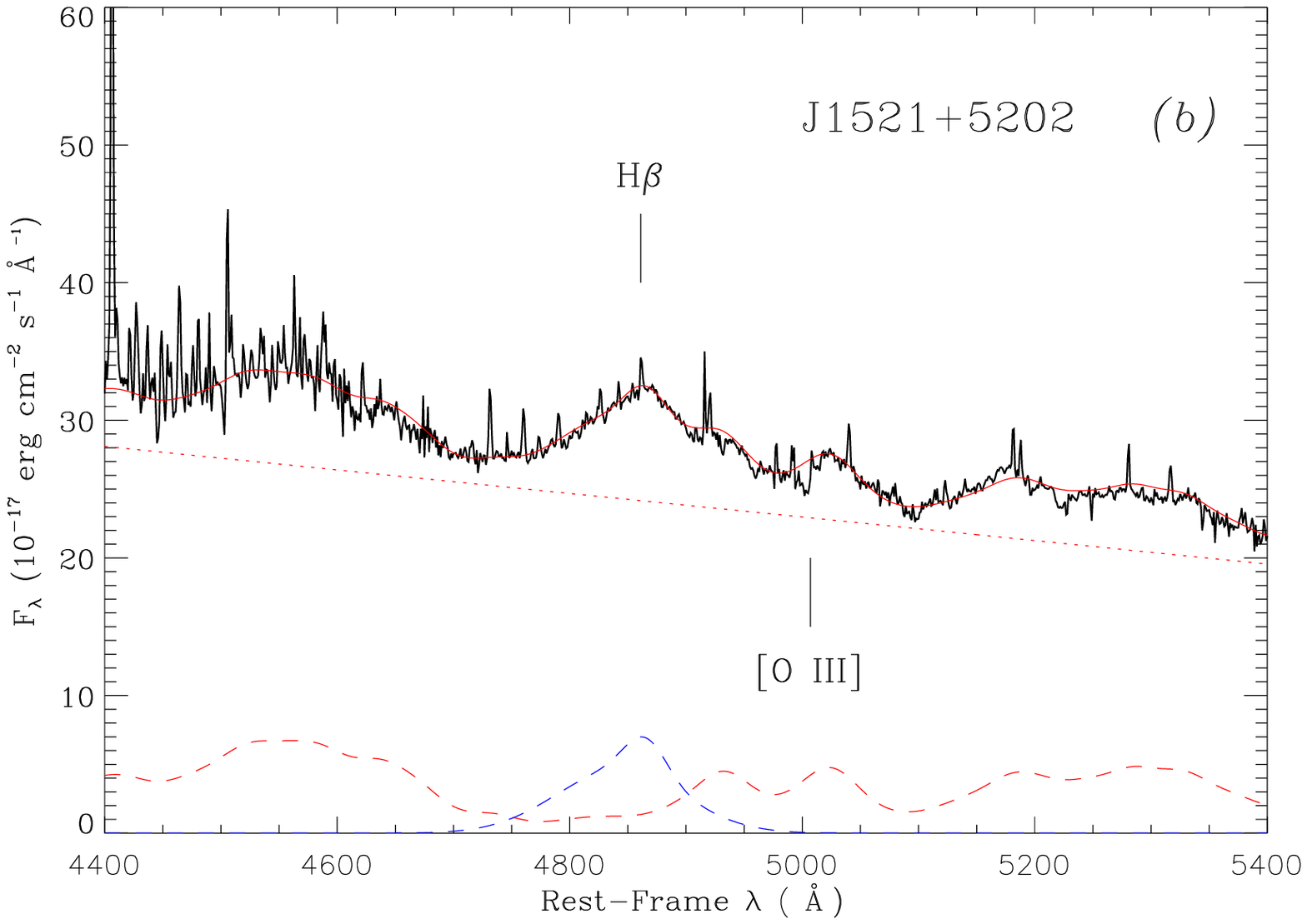}
    \caption{\footnotesize{$(a)$: The near-infrared spectrum of J1521+5202. The locations of the  
             H$\alpha$, H$\beta$, and [O~{\sc iii}] emission lines are labelled. The gaps within the 
             spectrum show the wavelength ranges that are affected most by the atmosphere.
             $(b)$: The modeling of the spectrum between 4400~\AA\ and 5400~\AA. The spectrum
             (black solid line) was resampled in bins of 1~\AA\ for clarity. The \hbox{best-fit} model
             (red solid line) consists of a continuum (red dotted line), a broadened Fe~{\sc ii}
             emission complex (red dashed line), and an H$\beta$ emission line (blue dashed line).
             The spectral resolution is $R\approx$~2500.}
             \label{j1521_fig}}
\end{figure*}
The \xray\ properties of J1521+5202 were first reported by Just et~al. (2007). This extremely 
optically luminous ($M_i=-30.19$) quasar is exceptionally \xray\ weak by a factor
of 34.5 (see \S\ref{xray}). It has unusual UV emission-line properties (see Fig.~\ref{spec_fig} 
and Table~\ref{qso_table}) like those of \phl, for which the \xray\ weakness may perhaps be attributed to 
a high $L/L_{\rm Edd}$ value
(Leighly et~al. 2007a). To investigate the cause of the \xray\ weakness for 
J1521+5202 and its relation to \phl, we obtained its \hbox{near-infrared} spectrum ($JHK$ bands) including coverage of 
H$\alpha$, H$\beta$, and [\ion{O}{3}]~$\lambda5007$.

The near-infrared spectrum for J1521+5202 is shown in Fig.~\ref{j1521_fig}$(a)$ with a spectral coverage
of 0.95--2.45~$\mu$m. We obtained this spectrum using the TripleSpec near-infrared spectrograph (Wilson et~al. 2004)
on the Apache Point Observatory (APO) 3.5 m telescope on 2009~May~19. We used a slit width of 1.7\arcsec, corresponding 
to a spectral resolution of $R\approx2500$.  This relatively broad slit width was chosen to maximize the \hbox{signal-to-noise} 
ratio in key 
spectral features such as the H${\beta}$ and [\ion{O}{3}]~$\lambda5007$ emission lines. We observed J1521+5202 
using an ``ABBA'' nodding pattern along the slit with each individual exposure lasting 200~s.  We collected twelve 
sets of observations in this way, giving $\approx 9600$~s of total exposure for the J1521+5202 spectrum.  Every 
$\approx40$ min, we observed a calibration star (HD$143817$) for 40 s. At the end of the observing 
run, we collected calibration flat frames. Sky lines in the images were used to perform wavelength calibration. 
We reduced the data following the standard procedures implemented in the 2009~May~1 version of the 
{\tt TripleSpecTool} package. {\tt TripleSpecTool} is a modified version of {\tt Spextool} (Vacca et~al. 2003; 
Cushing et~al. 2004) that generates \hbox{telluric-absorption} corrected, \hbox{absolute-calibrated} 
spectra from TripleSpec images. 

The near-infrared spectrum was modeled following the methods of Shemmer et~al. (2004), as summarized below. 
Our model consists of (1) a linear continuum fitted between two narrow ($\pm20$~\AA) bands
centered on 4700~\AA\ and 5100~\AA\ in the rest frame; (2) an \ion{Fe}{2} emission template (Boroson \& Green 1992)
broadened by convolution with a Gaussian profile with FWHM~$=2700$~km~s$^{-1}$; 
(3) the broad H$\beta$ component represented by a Gaussian profile with $1200$~km~s$^{-1}$ $\leq$ FWHM $\leq$ $15000$~km~s$^{-1}$;
and (4) the narrow H$\beta$ component and [\ion{O}{3}]~$\lambda\lambda$4959,~5007 represented by three Gaussian
profiles with $300$~km~s$^{-1}$~$\leq$~FWHM $\leq$~$1200$~km~s$^{-1}$. The three narrow Gaussian profiles were 
set to have the same width. The [\ion{O}{3}] lines have the theoretical ratio 
\hbox{$I$([\ion{O}{3}]~$\lambda5007$) / $I$([\ion{O}{3}]~$\lambda4959$) $=2.95$}. 
The best-fit model is shown in Fig.~\ref{j1521_fig}$(b)$. The [\ion{O}{3}] lines are undetected; 
we only obtained an upper limit of $W_r$([\ion{O}{3}]~$\lambda5007$) $<1$~\AA. 
To set this upper limit, we assumed a Gaussian profile for [\ion{O}{3}]~$\lambda5007$ with 
FWHM $=1000$~km~s$^{-1}$, and we then determined the weakest such feature that would have
been detected in our spectrum (e.g., Netzer et~al. 2004).

We measured the systemic redshift using the best-fit H$\beta$ line, $z=2.238$. This redshift is larger
than that determined from the \ion{Mg}{2} emission in the SDSS spectrum ($z_q=2.19$). Using the H$\beta$ redshift, 
the \ion{C}{4} line of J1521+5202 has an even larger blueshift ($-9400$~km~s$^{-1}$). 
The \ion{C}{4} blueshift remains strong ($-8400$~km~s$^{-1}$) using the redshift value estimated from H$\alpha$.
Like \phl, J1521+5202 has very weak [\ion{O}{3}]~$\lambda5007$ emission, and strong \ion{Fe}{2} emission, which
characterizes this source as an extreme eigenvector 1 object (Boroson \& Green 1992). The broad H$\beta$ line of
J1521+5202 has FWHM(H$\beta$) $=5750$~km~s$^{-1}$.
We also measured the rest-frame equivalent width of H$\beta$, $W_r$(H$\beta$) $=30.7$~\AA, which is similar to those of 
the two WLQs in Shemmer et~al. (2010), and smaller than that of \phl\ (50~\AA). Like WLQs, J1521+5202 has weaker 
H$\beta$ emission than most quasars (Boroson \& Green 1992; Shemmer et~al. 2010). The strength
of optical \ion{Fe}{2} emission was measured to be $W_r$(\ion{Fe}{2}) $=50.3$~\AA\ between 4434~\AA\ and 4684~\AA, yielding 
an \ion{Fe}{2}/H$\beta$ ratio of 1.6, which is similar to those for \phl\ and the two WLQs in Shemmer et~al. (2010), 
and larger than those for 
most typical quasars (e.g., Netzer et~al. 2004). This is expected given the weak [\ion{O}{3}] emission and the anti-correlation
between $W_r$(\ion{O}{3}) and the \ion{Fe}{2}/H$\beta$ ratio (see Fig.~9 of Netzer et~al. 2004). 

The virial $M_{\rm BH}$ and $L/L_{\rm Edd}$ for J1521+5202 have been estimated from its rest-frame optical properties, 
FWHM(H$\beta$) and $f_\lambda(5100$~\AA) ($2.20\times10^{-16}$~erg~cm$^{-2}$~s$^{-1}$~\AA$^{-1}$), using the
empirical relation between BELR size and luminosity in Kaspi et~al. (2005) modified by Bentz et~al. (2009), and
shown in Equations (1) and (2) of Shemmer et~al. (2010).
The bolometric correction to $\nu L_{\nu}(5100\,\mbox{\AA})$, $f(L)=5.5$, was calculated with Equation (21) 
of Marconi et~al. 
(2004). The results for J1521+5202 are $M_{\rm BH}=6.2\times10^9M_{\sun}$ and \hbox{$L/L_{\rm Edd}=0.81$}. The $L/L_{\rm Edd}$ value 
for J1521+5202 is not extraordinary compared to the \hbox{high-redshift} 
($2<z<3.5$), luminous ($L\gtrsim10^{46}$~erg~s$^{-1}$) quasars 
in Shemmer et~al. (2004), for which the $M_{\rm BH}$ and $L/L_{\rm Edd}$ are measured in a similar single-epoch spectroscopy approach. 
However, we recognize the considerable uncertainties of $M_{\rm BH}$ and $L/L_{\rm Edd}$ estimated with this approach
(at least a factor of 2--3 and perhaps much larger). When $L/L_{\rm Edd}$ is close to unity, radiation pressure becomes
important, which may present difficulties for the virial motion assumption for the gas around the SMBH 
(e.g., Marconi et~al. 2008, 2009; 
but also see Netzer \& Marziani 2010). Furthermore, the calibration of $M_{\rm BH}$ with FWHM(H$\beta$) and  
$L_\nu(5100$~\AA) is performed using reverberation-mapping data for low-luminosity, low-redshift ($z<0.3$) AGNs. 
Applying this calibration to 
high-luminosity, high-redshift AGNs may produce further uncertainties (e.g., Kaspi et~al. 2007; also see Footnote~24). 




\clearpage
\begin{deluxetable*}{clccccc}
\tablecolumns{7} \tabletypesize{\footnotesize}
\tablewidth{0pt}
\tablecaption{X-ray Observation Log\label{log_table}}
\tablehead{
    \colhead{} & \colhead{} & \colhead{} & \colhead{$\Delta_{{\rm Opt-X}}$\tablenotemark{b}}
               & \colhead{Observation} & \colhead{Observation} & \colhead{Exp. Time\tablenotemark{c}} \\
    \colhead{} & \colhead{Object Name (SDSS~J)} & \colhead{$z$\tablenotemark{a}}  & \colhead{(arcsec)}
               & \colhead{Date}  & \colhead{ID} & \colhead{(ks)}
}
\startdata
\multicolumn{2}{l}{{\it Chandra} Cycle 11 Objects} & & & & &\\
& $000009.38+135618.4$ & $2.234$ & $0.4$ & 2009 Aug 15 & 11490 & $9.5$ \\	
& $084842.63+540808.2$ & $2.225$ & \nodata & 2009 Dec 27 & 11492 & $13.1$ \\
& $090312.22+070832.4$ & $2.224$ & $0.5$ & 2009 Dec 29 & 11491 & $11.5$ \\
& $094602.31+274407.0$ & $2.382$ & $0.1$ & 2010 Jan 16 & 11489 & $5.0$ \\
& $112551.88+502803.6$ & $2.200$ & $0.1$ & 2010 Sep 22 & 11495 & $9.8$ \\
& $123035.46+332000.5$ & $2.287$ & \nodata & 2010 Jul 20 & 11493 & $5.3$ \\
& $145453.53+032456.8$ & $2.367$ & $0.0$ & 2009 Dec 19 & 11494 & $6.3$ \\
& $161801.71+070450.2$ & $2.235$ & $0.2$ & 2010 Jan 01 & 11496 & $7.7$ \\
\\
\multicolumn{2}{l}{Archival \xray\ Data Objects} & & & & &\\
& $121946.20+124454.1$ & $2.233$ & \nodata & 2008 Feb 14 & 8039 & $5.1$ \\
& $123042.52+204941.3$ & $2.280$ & \nodata & 2001 Jul 01 & 0112650101 & $23.9$\tablenotemark{d} \\
& $152156.48+520238.5$\tablenotemark{e} & $2.190$ & $0.1$ & 2006 Jul 16 & 6808 & $4.1$
\enddata
\tablenotetext{a}{Redshift for each source, adopted as the $z_{q}$ value in Table~\ref{abs_table}.}
\tablenotetext{b}{Angular distance between the optical and \xray\ positions; no entry indicates no \xray\ detection.}
\tablenotetext{c}{The {\it Chandra} and {\it XMM-Newton} exposure times are 
corrected for detector dead time.}
\tablenotetext{d}{The MOS exposure time is listed here. The exposure time for the pn detector was 17.5~ks.}
\tablenotetext{e}{This object was discussed in Just et~al. (2007).}
\end{deluxetable*}

\begin{deluxetable*}{ccccl}
\tabletypesize{\footnotesize}
\setlength{\tabcolsep}{0.04in}
\tablecaption{Quasar Redshifts and Narrow Absorption-Line Properties\label{abs_table}}
\tablewidth{0pt}
\tablehead{
\colhead{Object Name (SDSS~J)} & \colhead{$z_{CAS}$\tablenotemark{a}} & \colhead{$z_{q}$\tablenotemark{b}} & \colhead{$z_{HW}$\tablenotemark{c}} & \colhead{Redshifts of Absorption Systems}
}
\startdata
000009.38+135618.4 & 2.2342$\pm$0.0015 & 2.2342$\pm$0.0015 & 2.2405$\pm$0.0015 & 0.9635 \\
084842.63+540808.2 & 2.2252$\pm$0.0030 & 2.2252$\pm$0.0030 & 2.2065$\pm$0.0017 & 2.228\tablenotemark{d} \\
090312.22+070832.4 & 2.2237$\pm$0.0020 & 2.2237$\pm$0.0020 & 2.2145$\pm$0.0018 & 2.254\tablenotemark{d,}\tablenotemark{e}, 1.756 \\
094602.31+274407.0 & 2.3823$\pm$0.0007 & 2.3823$\pm$0.0007 & 2.3828$\pm$0.0007 & 2.323, 1.926, 1.029 \\
112551.88+502803.6 & 2.1758$\pm$0.0014 & 2.200$\pm$0.005 & 2.1778$\pm$0.0015 & 1.136, 1.0825 \\
121946.20+124454.1 & 2.1813$\pm$0.0027 & 2.2333$\pm$0.015 & 2.1497$\pm$0.0013 & 2.2333\tablenotemark{d}, 2.2086\tablenotemark{d}, 1.767, 1.631 \\
123035.46+332000.5 & 2.2865$\pm$0.0017 & 2.2865$\pm$0.0017 & 2.3057$\pm$0.0010 & 0.928, 0.7675 \\
123042.52+204941.3 & 2.2249$\pm$0.0018 & 2.280$\pm$0.015 & \nodata & 2.289\tablenotemark{d}, 2.28\tablenotemark{d}, 0.727, 0.618 \\
145453.53+032456.8 & 2.3672$\pm$0.0013\tablenotemark{f} & 2.3672$\pm$0.0013 & 2.3613$\pm$0.0023 & 2.302, 1.5245, 0.789 \\
152156.48+520238.5 & 2.208$\pm$0.011   & 2.190$\pm$0.011   & 2.1648$\pm$0.0017 & 1.875, 1.616 \\
161801.71+070450.2 & 2.2347$\pm$0.0019 & 2.2347$\pm$0.0019 & 2.2195$\pm$0.0021 & 2.1135, 2.075 
\enddata
\tablenotetext{a}{Redshifts from the SDSS CAS.}
\tablenotetext{b}{Adopted redshifts for the \phl\ analogs.}
\tablenotetext{c}{Redshifts from Hewett \& Wild (2010).}
\tablenotetext{d}{Associated narrow absorption lines (defined as $|v|<5000$~km~s$^{-1}$).}
\tablenotetext{e}{This system was only noticed in the HET spectrum (see \S\ref{sample:0903}).}
\tablenotetext{f}{Average for the CAS redshifts of the two independent spectra of this object: $z=2.3660\pm 0.0019$ (MJD 52029) and $z=2.3683\pm 0.0017$ (MJD 52045).}
\end{deluxetable*}

\clearpage
\begin{turnpage}
\begin{deluxetable}{ccccccrrrrc}
\tabletypesize{\scriptsize}
\tablecaption{Quasar UV Spectral Properties\label{qso_table}}
\tablewidth{0pt}
\tablehead{
\colhead{Object Name (SDSS~J)} & \colhead{MJD} & 
\colhead{C\,{\sc iv} Blueshift} & \colhead{C\,{\sc iv} FWHM} & \colhead{C\,{\sc iv} $\sigma_{\rm line}$} 
& \colhead{C~{\sc iv} FWHM/$\sigma_{\rm line}$}
& \colhead{$W_r$(C\,{\sc iv})} & \colhead{$W_r$(Si\,{\sc iv})\tablenotemark{a}} & \colhead{$W_r$($\lambda1900$~\AA)\tablenotemark{b}} & \colhead{$W_r$(Fe\,{\sc iii})} & \colhead{$\alpha_\nu$\tablenotemark{c}}
}
\startdata
000009.38+135618.4 &	52235 &	$ -5200 \pm 1000 $ &	$  6400 \pm 500 $ &	$ 4800 \pm  800 $ & $ 1.33 \pm 0.25 $ &	$  5.0 \pm 1.5 $ &	$ 2.8 \pm 1.5 $ &	$  4.4 \pm 1.8 $ &	$ 2.8 \pm 1.2 $ & $-0.97$ \\
084842.63+540808.2 &	51899 &	$ -4600 \pm  500 $ &	$  7000 \pm 500 $ &	$ 5600 \pm  800 $ & $ 1.25 \pm 0.20 $ &	$ 10.3 \pm 1.8 $ &	$ 4.4 \pm 1.2 $ &	$ 11.0 \pm 2.4 $ &	$ 5.0 \pm 1.8 $ & $-0.50$ \\
090312.22+070832.4 &	52674 &	$ -2300 \pm  600 $ &	$  6900 \pm 600 $ &	$ 2900 \pm  500 $ & $ 2.38 \pm 0.46 $ &	$  6.5 \pm 1.8 $ &	$ 2.9 \pm 1.5 $ &	$  8.5 \pm 2.4 $ &	$ 6.2 \pm 1.8 $ & $-0.49$ \\
090312.22+070832.4 (HET) & 55506 & $  -610 \pm  300 $ &    $  6600 \pm 690 $ &     $ 3240 \pm  680 $ & $ 2.03 \pm 0.45 $ & $  5.2 \pm 0.7 $ &      $ 3.8 \pm 0.8 $ &       $  9.5 \pm 0.7 $ &      $ 6.2 \pm 0.7 $ & \nodata \\
094602.31+274407.0 &	53385 &	$ -1700 \pm  500 $ &	$ 11000 \pm 800 $ &	$ 5800 \pm  800 $ & $ 1.90 \pm 0.30 $ &	$  6.0 \pm 0.9 $ &	$ 4.2 \pm 0.6 $ &	$  7.3 \pm 0.9 $ &	$ 6.8 \pm 0.9 $ & $-0.68$ \\
112551.88+502803.6 &	52365 &	$ -3400 \pm  700 $ &	$  7400 \pm 600 $ &	$ 3500 \pm  600 $ & $ 2.11 \pm 0.40 $ &	$  7.3 \pm 1.8 $ &	$ 6.9 \pm 1.8 $ &	$  8.8 \pm 2.1 $ &	$ 3.8 \pm 1.5 $ & $-0.53$ \\
121946.20+124454.1 &	53115 &	$ -9400 \pm  300 $ &	$  7100 \pm 500 $ &	$ 7500 \pm 1000 $ & $ 0.95 \pm 0.14 $ &	$  8.1 \pm 0.6 $ &	$ 3.4 \pm 0.6 $ &	$  8.3 \pm 0.9 $ &	$ 7.7 \pm 0.6 $ & $-0.87$ \\
123035.46+332000.5 &	53472 &	$ -2800 \pm  600 $ &	$ 11000 \pm 800 $ &	$ 5600 \pm  800 $ & $ 1.96 \pm 0.31 $ &	$ 10.4 \pm 1.2 $ &	$ 4.0 \pm 0.9 $ &	$ 12.2 \pm 1.2 $ &	$ 6.8 \pm 1.2 $ & $-0.32$ \\
123042.52+204941.3 &	54480 &	$ -5700 \pm  800 $ &	$  4900 \pm 300 $ &	$ 8600 \pm 1200 $ & $ 0.57 \pm 0.09 $ &	$  6.8 \pm 1.5 $ &	$ 2.2 \pm 1.2 $ &	$  3.5 \pm 1.5 $ &	$ 5.6 \pm 1.2 $ & $-0.88$ \\
145453.53+032456.8\tablenotemark{d} &	52037 &	$ -1600 \pm  600 $ &	$  5500 \pm 500 $ &	$ 3900 \pm  600 $ & $ 1.41 \pm 0.25 $ &	$  5.1 \pm 1.3 $ &	$ 3.2 \pm 2.1 $ &	$  5.2 \pm 1.2 $ &	$ 7.2 \pm 1.4 $ & $-0.28$ \\
152156.48+520238.5 &	52376 &	$ -4900 \pm  300 $ &	$ 11700 \pm 800 $ &	$ 5700 \pm  800 $ & $ 2.05 \pm 0.32 $ &	$  9.1 \pm 0.6 $ &	$ 2.7 \pm 0.3 $ &	$  8.2 \pm 0.6 $ &	$ 7.1 \pm 0.6 $ & $-0.61$ \\
161801.71+070450.2 &	53884 &	$  -800 \pm  600 $ &	$  7700 \pm 500 $ &	$ 5900 \pm  800 $ & $ 1.31 \pm 0.20 $ &	$  8.3 \pm 1.5 $ &	$ 3.5 \pm 1.2 $ &	$ 10.1 \pm 1.5 $ &	$ 3.9 \pm 1.2 $ & $-0.38$ \\
PHL~1811	& \nodata &	$ -1400 \pm  250 $ &	$  4300 \pm 700 $ &	$ 3400 \pm 1100 $ & $ 1.26 \pm 0.46 $ &	$  4.7 \pm 0.9 $ &	$ 4.8 \pm 0.9 $ &	$  8.3 \pm 0.6 $ &	$ 4.7 \pm 0.6 $ & $-0.74$ \\
Average \tablenotemark{e}	 & \nodata &	$ -3650 \pm 2450 $ &	$ 7600 \pm 2400 $ & $ 5300 \pm 1700 $ & $ 1.44 \pm 0.65 $ &	$  7.3 \pm 2.0 $ &	$ 3.8 \pm 1.3 $ &	$  8.0 \pm 2.6 $ &	$ 5.6 \pm 1.6 $ & $-0.60$ \\ \\
V01~composite\ \tablenotemark{f}	& \nodata &	$  -570 \pm   30 $ &	$  5050 \pm 550 $ &	$ 4500 \pm 1000 $ & $ 1.12 \pm 0.28 $ &	$ 30.0 \pm 0.3 $ &	$ 8.7 \pm 0.3 $ &	$ 21.7 \pm 0.2 $ &	$ 2.9 \pm 0.1 $ & $-0.44$ \\
Y04~composite~B4\ \tablenotemark{g} & \nodata & $ -470 \pm  260 $ &	$  4380 \pm 250 $ &	$ 4250 \pm  240 $ & $ 1.03 \pm 0.08 $ &	$ 26.0 \pm 0.3 $ &	$ 8.2 \pm 0.3 $ &	$ 22.4 \pm 0.2 $ &	$ 3.1 \pm 0.1 $ & $-0.42$ \\ \\
PHL~1092\ \tablenotemark{h} & \nodata & $ -3900 \pm  500 $ &	$  8740 \pm 1080 $ &	$ 2690 \pm  700 $ & $ 3.25 \pm 0.29 $ &	$ 7.2 \pm 1.2 $ &	$ 3.9 \pm 0.9 $ &	$ 15.4 \pm 1.2 $ &	$ 3.3 \pm 1.2 $ & $-0.70$
\enddata
\tablecomments{All blueshift, FWHM, and $\sigma_{\rm line}$ values are in units of km s$^{-1}$.
All $W_r$ values are in units of \AA.
The uncertainties for the C\,{\sc iv} FWHM are two times the formal $\sigma$ values,
the uncertainties for the C\,{\sc iv} $\sigma_{\rm line}$ are four times the formal $\sigma$ values,
and the uncertainties for the $W_r$ values are three times the formal $\sigma$ values.
These scalings were estimated using the two independent spectra of J1454+0324; 
each scaling is the rounded ratio of the RMS of the two independent measurements 
and the average of their formal $\sigma$ values.}
\tablenotetext{a}{This line is a blend of Si~{\sc iv} and O~{\sc iv}]; we refer to it as Si~{\sc iv}
simply for convenience.}
\tablenotetext{b}{Mainly C\,{\sc iii}] $\lambda 1909$, but also including [Ne\,{\sc iii}] $\lambda 1814$, 
Si\,{\sc ii} $\lambda 1816$, Al\,{\sc iii} $\lambda 1857$, Si\,{\sc iii}] $\lambda 1892$, 
and several Fe\,{\sc iii} multiplets (see Table 2 of Vanden Berk et al. 2001).}
\tablenotetext{c}{The spectral index of the presumed \pl\ continuum, where \hbox{$f_\nu \propto \nu^{\alpha_\nu}$}.}
\tablenotetext{d}{All the spectral parameters for J1454+0324, including MJD, are the average of two spectra taken at MJD=52029 and MJD=52045.}
\tablenotetext{e}{The average values for the \phl\ analogs plus \phl\ itself. }
\tablenotetext{f}{The composite spectrum from Vanden Berk et al. (2001).}
\tablenotetext{g}{The `B4' composite spectrum from Yip et al. (2004).}
\tablenotetext{h}{We measured the emission-line properties for PHL~1092 using its {\it Hubble Space Telescope} STIS spectrum (Leighly et~al. 2007b). See Footnote~20 for further discussion.}
\end{deluxetable}
\end{turnpage}

\clearpage
\begin{center}
\begin{deluxetable}{llccccc}
\tablecolumns{7} \tabletypesize{\footnotesize}
\tablewidth{0pt}
\tablecaption{X-ray Counts\label{cts_table}}
\tablehead{
 \colhead{} &    
    \colhead{}                     
    & \colhead{Full Band}     
    & \colhead{Soft Band}     
    & \colhead{Hard Band}     
    & \colhead{Band}
    & \colhead{}\\
 \colhead{} &
    \colhead{Object Name (SDSS J)} 
    & \colhead{(0.5--8.0 keV)\tablenotemark{a}}
    & \colhead{(0.5--2.0 keV)\tablenotemark{a}} 
    & \colhead{(2.0--8.0 keV)\tablenotemark{a}} 
    & \colhead{Ratio\tablenotemark{b}}  
    & \colhead{$\Gamma$\tablenotemark{c}}
}
\startdata
\multicolumn{2}{l}{{\it Chandra} Cycle 11 Objects} & & & \\
& $000009.38+135618.4$ & $4.8^{+3.4}_{-2.1}$ & $3.8^{+3.1}_{-1.9}$ & $<5.1$ & $<1.24$ & $>0.58$ \\
& $084842.63+540808.2$ & $<6.5$ & $<4.8$ & $<5.1$ & \nodata & \nodata \\
& $090312.22+070832.4$ & $24.8^{+6.1}_{-4.9}$ & $22.7^{+5.8}_{-4.7}$ & $<7.0$ & $<0.31$ & $>1.86$ \\
& $094602.31+274407.0$ & $5.1^{+3.4}_{-2.2}$ & $4.0^{+3.2}_{-1.9}$ & $<5.2$ & $<1.23$ & $>0.55$ \\
& $112551.88+502803.6$ & $6.9^{+3.8}_{-2.6}$ & $4.0^{+3.2}_{-1.9}$ & $3.2^{+3.0}_{-1.7}$ & $0.78^{+1.02}_{-0.54} $ & $0.96^{+1.04}_{-0.75} $ \\
& $123035.46+332000.5$ & $<5.2$ & $<3.2$ & $<5.2$ & \nodata & \nodata \\
& $145453.53+032456.8$ & $413.7^{+21.4}_{-20.3}$ & $338.6^{+19.4}_{-18.4}$ & $78.7^{+9.9}_{-8.9}$ & $0.23^{+0.03}_{-0.03} $ & $2.09^{+0.13}_{-0.11} $ \\
& $161801.71+070450.2$ & $190.3^{+14.8}_{-13.8}$ & $134.1^{+12.6}_{-11.6}$ & $56.2^{+8.5}_{-7.5}$ & $0.44^{+0.08}_{-0.07} $ & $1.53^{+0.16}_{-0.15} $ \\

\\
 \multicolumn{2}{l}{Archival \xray\ Data Objects} & & & \\
& $121946.20+124454.1$ & $<5.8$ & $<5.0$ & $<4.1$ & \nodata & \nodata \\
& $123042.52+204941.3$ & $<15.5$ & $<12.2$ & $<11.7$ & \nodata & \nodata \\
& $152156.48+520238.5$ & $3.2^{+3.0}_{-1.7}$ & $<5.2$ & $2.2^{+2.7}_{-1.4}$ & $>0.44$ & $<1.40$ 
\enddata
\tablenotetext{a}{Errors on the \hbox{X-ray} counts were calculated using Poisson statistics corresponding to the 1$\sigma$ significance level according to Tables 1 and 2 of Gehrels (1986).}
\tablenotetext{b}{The band ratio is defined here as the number of hard-band counts divided by the number of soft-band counts. The errors on the band ratio correspond to the 1$\sigma$ significance level and were calculated using equation (1.31) in \S 1.7.3 of Lyons (1991). The band ratios for all of the {\it Chandra} objects observed in the same cycle can be directly compared with one another.}
\tablenotetext{c}{The effective power-law photon indices were calculated using the {\it Chandra} PIMMS tool (version 3.9$k$). The effect of the quantum efficiency decay over time at low energies of the ACIS detector were corrected for {\it Chandra} observed objects.}
\end{deluxetable}
\end{center}

\clearpage
\begin{turnpage}
\begin{deluxetable}{ccccccccccccccc}
\tablecolumns{15} \tabletypesize{\scriptsize}
\tablewidth{0pt}
\tablecaption{X-ray and Optical Properties\label{aox_table}}
\tablehead{
    \colhead{} &
    \colhead{}                     
    & \colhead{}        
    & \colhead{}        
    & \colhead{}            
    & \colhead{Count}
    & \colhead{}                       
    & \colhead{}                 
    & \colhead{log $L$}                 
    & \colhead{}
    & \colhead{log $L_{\nu}$}          
    & \colhead{} 
    & \colhead{}
    & \colhead{}
    & \colhead{} \\
    \colhead{} &
    \colhead{Object Name (SDSS J)} 
    & \colhead{$m_{i}$}
    & \colhead{$M_{i}$} 
    & \colhead{$N_{\rm H}$} 
    & \colhead{Rate\tablenotemark{a}} 
    & \colhead{$F_{0.5-2\;{\rm keV}}$\tablenotemark{b}} 
    & \colhead{$f_{\rm 2\;keV}$}
    & \colhead{($2-10\;{\rm keV}$)} 
    & \colhead{$f_{2500\mbox{\rm~\scriptsize\AA}}$}
    & \colhead{(2500 \AA)}  
    & \colhead{$\alpha_{\rm ox}$} 
    & \colhead{$\Delta \alpha_{\rm ox}$ $(\sigma)$\tablenotemark{c}}
    & \colhead{$f_{\rm x-weak}$\tablenotemark{d}}
    & \colhead{$R$} \\
    \colhead{} &
    \colhead{(1)} 
    & \colhead{(2)} 
    & \colhead{(3)} 
    & \colhead{(4)} 
    & \colhead{(5)} 
    & \colhead{(6)} 
    & \colhead{(7)} 
    & \colhead{(8)} 
    & \colhead{(9)} 
    & \colhead{(10)} 
    & \colhead{(11)} 
    & \colhead{(12)} 
    & \colhead{(13)} 
    & \colhead{(14)}
}
\startdata
\multicolumn{2}{l}{{\it Chandra} Cycle 11 Objects} & & & \\
& $000009.38+135618.4$ & $18.30$ & $-27.44$ & $3.77$ & $0.40^{+0.33}_{-0.19}$ & $0.17$ & $0.83$ & $43.90$ & $2.02$ & $31.39$ & $-2.07$ & $-0.38$ $(2.58)$ & $9.77$ & $<4.7$ \\
& $084842.63+540808.2$ & $18.76$ & $-26.90$ & $3.09$ & $<0.37$ & $<0.16$ & $<0.76$ & $<43.85$ & $1.12$ & $31.13$ & $<-1.98$ & $<-0.33$ $(2.27)$ & $>7.24$ & $<2.6$ \\
& $090312.22+070832.4$ & $18.64$ & $-27.11$ & $4.54$ & $1.98^{+0.51}_{-0.41}$ & $0.87$ & $4.21$ & $44.59$ & $1.45$ & $31.24$ & $-1.74$ & $-0.07$ $(0.50)$ & $1.52$ & $<2.4$ \\
& $094602.31+274407.0$ & $17.00$ & $-28.81$ & $1.73$ & $0.80^{+0.64}_{-0.38}$ & $0.33$ & $1.65$ & $44.24$ & $6.57$ & $31.95$ & $-2.15$ & $-0.38$ $(2.61)$ & $9.77$ & $4.0$ \\
& $112551.88+520803.6$ & $18.45$ & $-27.13$ & $1.25$ & $0.41^{+0.33}_{-0.20}$ & $0.16$ & $0.79$ & $43.86$ & $1.62$ & $31.28$ & $-2.04$ & $-0.36$ $(2.50)$ & $8.67$ & $<1.2$ \\
& $123035.46+332000.5$ & $17.80$ & $-27.90$ & $1.40$ & $<0.60$ & $<0.24$ & $<1.19$ & $<44.07$ & $2.70$ & $31.53$ & $<-2.06$ & $<-0.35$ $(2.37)$ & $>8.16$ & $<1.0$ \\
& $145453.53+032456.8$ & $17.95$ & $-27.88$ & $3.59$ & $53.36^{+3.06}_{-2.90}$ & $22.97$ & $115.39$ & $46.08$ & $2.75$ & $31.57$ & $-1.30$ & $0.42$ $(2.86)$ & $0.08$ & $12.8$ \\
& $161801.71+070450.2$ & $18.17$ & $-27.58$ & $4.59$ & $17.47^{+1.64}_{-1.51}$ & $7.74$ & $37.37$ & $45.55$ & $1.62$ & $31.29$ & $-1.40$ & $0.28$ $(1.92)$ & $0.19$ & $35.0$ \\
\\
\multicolumn{2}{l}{Archival \xray\ Data Objects} & & & \\
& $121946.20+124454.1$ & $17.90$ & $-27.75$ & $2.65$ & $<0.97$ & $<0.37$ & $<1.78$ & $<44.23$ & $3.02$ & $31.56$ & $<-2.01$ & $<-0.29$ $(2.01)$ & $>5.70$ & $3.9$ \\
& $123042.52+204941.3$ & $18.45$ & $-27.22$ & $1.99$ & $<0.35$ & $<0.27$ & $<1.34$ & $<44.12$ & $1.39$ & $31.24$ & $<-1.93$ & $<-0.26$ $(1.76)$ & $>4.76$ & $<1.8$ \\
& $152156.48+520238.5$ & $15.44$ & $-30.19$ & $1.58$ & $0.78^{+0.73}_{-0.41}$ & $0.25$ & $1.18$ & $44.03$ & $26.11$ & $32.48$ & $-2.44$ & $-0.59$ $(4.53)$ & $34.46$ & $<0.1$ 
\enddata
\tablecomments{The detailed explanation of each column is given in \S\ref{xray}.}
\tablenotetext{a}{The count rate in the observed-frame soft \hbox{X-ray} band ($0.5$--$2.0$ keV) in units of $10^{-3}$ ${\rm s}^{-1}$, except for J1521+5202 where the count rate is in the observed-frame full band ($0.5$--$8.0$ keV) because this source is undetected in the soft band, but detected in the full band.}
\tablenotetext{b}{The Galactic absorption-corrected observed-frame flux between $0.5$--$2.0$ keV in units of $10^{-14}$ erg cm$^{-2}$ s$^{-1}$. The value for J1521+5202 is also for the flux between $0.5$--$2.0$ keV, calculated from the full-band count rate.}
\tablenotetext{c}{$\Delta\alpha_{\rm ox}$: the difference between the measured $\alpha_{\rm ox}$ and the expected $\alpha_{\rm ox}$, defined by the $\alpha_{\rm ox}-L_{2500~{\rm \AA}}$ relation in equation (3) of Just et al.~(2007); the statistical significance of this difference, $\sigma$, is measured in units of the RMS for $\alpha_{\rm ox}$ defined in Table 5 of Steffen et al.~(2006).}
\tablenotetext{d}{The factor of X-ray weakness compared to a typical quasar with similar optical/UV luminosity; see \S3.}
\end{deluxetable}
\end{turnpage}

\begin{deluxetable}{ccccccc}
\tabletypesize{\footnotesize}
\tablecaption{Results of Two-Sample Tests\label{twost_table}}
\tablewidth{0pt}
\tablehead{
\colhead{} & & \multicolumn{2}{c}{J0903+0708 excluded} & & \multicolumn{2}{c}{J0903+0708 included} \\
\cline{3-4}\cline{6-7} \\
\colhead{Method} & & \colhead{Statistic} & \colhead{Null-hypothesis Probability} & & \colhead{Statistic} & \colhead{Null-hypothesis Probability}
}
\startdata
Peto-Prentice & & $11.115$ & $1.06\times10^{-28}$ & & $8.779$ & $1.65\times10^{-18}$ \\
Gehan (Permutation Variance) & & $4.663$ & $3.12\times10^{-6}$ & & $4.702$ & $2.58\times10^{-6}$ \\
Gehan (Hypergeometric Variance) & & $3.959$  & $7.53\times10^{-5}$ & & $4.008$ & $6.12\times10^{-5}$ \\
logrank & & $5.415$ & $6.13\times10^{-8}$ & & $5.334$ & $9.61\times10^{-8}$ \\
Peto-Peto & & $4.687$ & $2.77\times10^{-6}$ & & $4.724$ & $2.31\times10^{-6}$
\enddata
\tablecomments{For the detailed definition of each test statistic, see Feigelson \& Nelson (1985) and references therein; also see the ASURV manual at http://astrostatistics.psu.edu/statcodes/asurv. The null-hypothesis probability was calculated from each test statistic using a Gaussian distribution, e.g., for Peto-Prentice test, $1-P_G=1.06\times10^{-28}$, where $P_G$ is the cumulative Gaussian probability at $11.115\sigma$.}
\end{deluxetable}

\begin{deluxetable}{cccccccc}
\tablecolumns{8} \tablenum{8} \tabletypesize{\footnotesize}
\tablewidth{0pt}
\tablecaption{X-ray Spectral Analysis}
\tablehead{
\colhead{}
    & \colhead{}
    & \multicolumn{2}{c}{Power Law}
    & \colhead{}
    & \multicolumn{3}{c}{Power Law}\\
    \colhead{}
    & \colhead{}
    & \multicolumn{2}{c}{with Galactic Absorption}
    & \colhead{}
    & \multicolumn{3}{c}{with Galactic and Intrinsic Absorption}
    \\
    \cline{3-4}\cline{6-8}\\    
\colhead{Object Name (SDSS~J)} 
    & \colhead{}
    & \colhead{$\Gamma$} 
    & \colhead{$\chi^2/\nu$} 
    & \colhead{}
    & \colhead{$\Gamma$} 
    & \colhead{$N_H (10^{22}{\rm cm}^{-2})$}
    & \colhead{$\chi^2/\nu$}
}
\startdata
145453.53+032456.8 & & $2.13^{+0.18}_{-0.17}$ & $10.25/16$ & & $2.13^{+0.29}_{-0.16}$ & $<1.36$ & $10.25/15$ \\
161801.71+070450.2 & & $1.48^{+0.29}_{-0.28}$ & $18.70/14$ & & $1.48^{+0.41}_{-0.27}$ & $<1.98$ & $18.70/13$ 
\enddata
\label{fit_table}
\end{deluxetable}

\clearpage
\begin{turnpage}
\begin{deluxetable}{ccrrrrrrrrrrrrrcrrc}
\tabletypesize{\footnotesize}
\tablecaption{Correlation Probabilities from Kendall's $\tau$ Tests and Spearman Rank-order Analyses\label{corr_table}}
\tablewidth{0pt} \tablenum{7}
\tablehead{
     \colhead{} & & \multicolumn{5}{c}{RQQ\tablenotemark{a} ($N=8$)\tablenotemark{d}} & & \multicolumn{5}{c}{RQQ+RIQ\tablenotemark{b} ($N=10$)\tablenotemark{d}} & & \multicolumn{5}{c}{RQQ+G08a\tablenotemark{c}+PHL~1811 ($N=141$)\tablenotemark{d}} \\
     \cline{3-7}\cline{9-13}\cline{15-19}\\
     \colhead{} & & \multicolumn{2}{c}{Kendall} & &\multicolumn{2}{c}{Spearman} & & \multicolumn{2}{c}{Kendall} & & \multicolumn{2}{c}{Spearman} & & \multicolumn{2}{c}{Kendall} & & \multicolumn{2}{c}{Spearman} \\
     \cline{3-4}\cline{6-7}\cline{9-10}\cline{12-13}\cline{15-16}\cline{18-19}\\
     \colhead{Correlation with \daox} & & \colhead{$\tau$} & \colhead{$1-P_{\tau}$} & & \colhead{$r_S$} & \colhead{$1-P_S$} & & \colhead{$\tau$} & \colhead{$1-P_{\tau}$} & & \colhead{$r_S$} & \colhead{$1-P_S$} & & \colhead{$\tau$} & \colhead{$1-P_{\tau}$} & & \colhead{$r_S$} & \colhead{$1-P_S$}
}
\startdata
C {\sc iv} Blueshift & & $0.36$ & $28.2\%$ & & $0.31$ & $58.9\%$ & & $1.95$ & $94.9\%$ & & $0.69$ & $96.1\%$ & & $4.77$ & $>99.99\%$ & & $0.39$ & $>99.99\%$ \\
C {\sc iv} FWHM & & $1.10$ & $72.7\%$ & & $-0.19$ & $38.7\%$ & & $0.98$ & $67.2\%$ & & $-0.28$ & $60.6\%$ & & \nodata & \nodata & & \nodata & \nodata \\
C {\sc iv} $\sigma_{\rm line}$ & & $1.10$ & $72.7\%$ & & $-0.56$ & $85.8\%$ & & $0.73$ & $53.7\%$ & & $-0.41$ & $77.5\%$ & & \nodata & \nodata & & \nodata & \nodata \\
C {\sc iv} FWHM/$\sigma_{\rm line}$ & & $0.36$ & $28.2\%$ & & $0.31$ & $58.9\%$ & & $0.00$ & $0.0\%$ & & $0.11$ & $25.0\%$ & & \nodata & \nodata & & \nodata & \nodata \\
$W_r$(C {\sc iv}) & & $0.36$ & $28.2\%$ & & $-0.38$ & $68.5\%$ & & $0.73$ & $53.5\%$ & & $-0.39$ & $75.7\%$ & & $6.04$ & $>99.99\%$ & & $0.49$ & $>99.99\%$ \\
$W_r$(Si {\sc iv}) & & $1.81$ & $92.9\%$ & & $0.66$ & $91.7\%$ & & $0.24$ & $19.3\%$ & & $0.26$ & $56.8\%$ & & \nodata & \nodata & & \nodata & \nodata \\
$W_r$($\lambda1900$~\AA) & & $0.36$ & $28.2\%$ & & $0.10$ & $21.6\%$ & & $0.00$ & $0.0\%$ & & $-0.00$ & $1.7\%$ & & $1.20$ & $76.88\%$ & & $0.11$ & $80.32\%$ \\
$W_r$(Fe {\sc iii}) & & $1.10$ & $72.7\%$ & & $-0.61$ & $89.2\%$ & & $0.00$ & $0.0\%$ & & $-0.20$ & $44.9\%$ & & \nodata & \nodata & & \nodata & \nodata
\enddata
\tablenotetext{a}{Radio-quiet \phl\ analogs in our sample.}
\tablenotetext{b}{Radio-quiet and radio-intermediate \phl\ analogs in our sample.}
\tablenotetext{c}{Radio-quiet, non-BAL quasars in Sample~B of Gibson et~al. (2008a).}
\tablenotetext{d}{$N$ is the sample size. The Kendall's $\tau$ test is preferred when $N\leq30$, otherwise the Spearman rank-order
analysis is preferred.}
\end{deluxetable}
\end{turnpage}


\end{document}